\documentclass[11pt]{article}

\usepackage{epsfig}
\usepackage{multirow}
\usepackage{amsmath}
\usepackage{amssymb}
\usepackage{comment}
\usepackage{fullpage}
\usepackage{bigstrut}
\usepackage{subfigure}
\usepackage{color}

\newcommand{\flrbk}[1]{b^{\lfloor \log_b k\rfloor #1}}

\newcommand{\buzz}[1]{\emph{#1}}

\newtheorem{theorem}{Theorem}
\newtheorem{lemma}[theorem]{Lemma}

\newtheorem{conjecture}[theorem]{Conjecture}

\newcommand{\prob}[1]{{\rm Prob}[#1]}

\newcommand{\newloglike}[2]{\newcommand{#1}{\mathop{\rm #2}\nolimits}}
\newloglike{\E}{E}
\newloglike{\sgn}{sgn}

\newcommand{\calF}{{\cal F}}

\newcommand{\etal}[1]{{\it et al.\/}}

\newenvironment{proof}{\noindent\par{\bf Proof: }}{\nopagebreak\rule{1 ex}{0.8 em}\medskip} 

\newcommand{\ceil}[1]{\left\lceil{#1}\right\rceil}
\newcommand{\floor}[1]{\left\lfloor{#1}\right\rfloor}

\begin{document}

\title{Fault-tolerant Routing in Peer-to-peer Systems\footnote{This
is an extended version of the paper appearing in the proceedings of the
\emph{Twenty-First ACM Symposium on Principles of Distributed Computing}, 
2002}}

\author{James Aspnes\thanks{
Department of Computer Science, Yale University,
New Haven, CT 06520-8285, USA.
Email: {\tt aspnes@cs.yale.edu}.
Supported by NSF grants CCR-9820888 and CCR-0098078.}
\and Zo\"{e} Diamadi\thanks{
Department of Computer Science, Yale University,
New Haven, CT 06520-8285, USA.
Email: {\tt diamadi@cs.yale.edu}.
Supported in part by ONR grant N00014-01-1-0795.}
\and Gauri Shah\thanks{
Department of Computer Science, Yale University,
New Haven, CT 06520-8285, USA.
Email: {\tt gauri.shah@yale.edu}.
Supported by NSF grants CCR-9820888 and CCR-0098078.}
}

\maketitle

\begin{abstract}
We consider the problem of designing an overlay network and routing
mechanism that permits finding resources efficiently in a peer-to-peer
system. We argue that many existing approaches to this problem can be
modeled as the construction of a random graph embedded in a metric 
space whose points represent resource identifiers, where the 
probability of a connection between two nodes depends only on the 
distance between them in the metric space.  We study the performance of 
a peer-to-peer system where nodes are embedded at grid points in a simple 
metric space: a one-dimensional real line. We prove upper and lower bounds 
on the message complexity of locating particular resources in such a system, 
under a variety of assumptions about failures of either nodes or the 
connections between them. Our lower bounds in particular show that the 
use of inverse power-law distributions in routing, as suggested by
Kleinberg~\cite{KL99}, is close to optimal. We also give efficient 
heuristics to dynamically maintain such a system as new nodes arrive and 
old nodes depart. Finally, we give experimental results that suggest 
promising directions for future work.
\end{abstract}

\section{Introduction}
\label{sec:INTRODCUTION}

Peer-to-peer systems are distributed systems without any central
authority and with varying computational power at each machine.
We study the problem of locating resources in such a large network
of heterogeneous machines that are subject to crash failures. We
describe how to construct distributed data structures that have
certain desirable properties and allow efficient resource location.

Decentralization is a critical feature of such a system
as any central server not only provides a vulnerable point of
failure but also wastes the power of the clients. Equally important
is scalability: the cost borne by each node must not depend too much
on the network size and should ideally be proportional, within
polylogarithmic factors, to the amount of data the node seeks or
provides. Since we expect nodes to arrive and depart at a high rate,
the system should be resilient to both link and node failures.
Furthermore, disruptions to parts of the data structure should
self-heal to provide self-stabilization.

Our approach provides a hash table-like functionality, based on 
keys that uniquely identify the system resources. To accomplish this, 
we map resources to points in a metric space either directly from 
their keys or from the keys' hash values. This mapping dictates an 
assignment of nodes to metric-space points. We construct and maintain 
a random graph linking these points and use greedy routing to 
traverse its edges to find data items. The principle we
rely on is that failures leave behind yet another (smaller) random 
graph, ensuring that the system is robust even in the face of 
considerable damage. Another compelling advantage of random graphs 
is that they eliminate the need for global coordination. Thus, we 
get a fully-distributed, egalitarian, scalable system with no 
bottlenecks.

We measure performance in terms of the number of messages sent by 
the system for a search or an insert operation. The self-repair 
mechanism may generate additional traffic, but we expect to amortize 
these costs over the search and insert operations.  Given the growing 
storage capacity of machines, we are less concerned with minimizing 
the storage at each node; but in any case
the space requirements are small. The 
information stored at a node consists only of a network address for
each neighbor.

The rest of the paper is organized as follows. 
Section~\ref{sec:APPROACH} explains our abstract model in detail,
and Section~\ref{sec:RELATED} describes some existing 
peer-to-peer systems. We prove our results for routing in 
Section~\ref{sec:ROUTING}. In Section~\ref{sec:RANDOMGRAPHS}, 
we present a heuristic method for constructing the random graph and
provide experimental results that show its performance in practice.
Section~\ref{sec:EXPERIMENTS} describes results of experiments
we performed to test the routing performance of our constructed 
distributed data structure. Conclusions and future work 
are discussed in Section~\ref{sec:CONCLUSIONS}.

\section{Our approach}
\label{sec:APPROACH}

\begin{figure}
   \centerline{\epsfig{figure=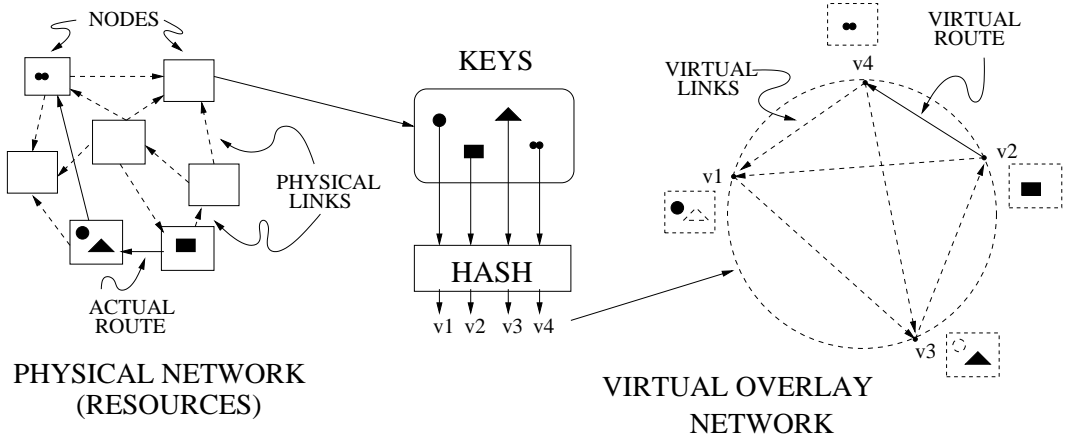, width=400pt}}
   \caption{An example of the metric-space embedding.}
   \label{mapping}
\end{figure}

The idea underlying our approach consists of three basic parts:
(1) embed resources as points in a metric space, (2) construct a
random graph by appropriately linking these points, and
(3) efficiently locate resources by routing greedily along the
edges of the graph. Let $R$ be a set of resources spread over a large,
heterogeneous network $N$. For each resource $r \in R$,
$owner(r)$ denotes the node in $N$ that provides $r$ and
$key(r)$ denotes the resource's key. Let $K$ be the set of all
possible keys.
We assume a hash function $h: K \rightarrow V$ such that
resource $r$ maps to the point
$v=h(key(r))$ in a metric space $(V,d)$, where $V$ is the point set
and $d$ is the distance metric as shown in Figure~\ref{mapping}.
The hash function is assumed to populate the metric space evenly.
Note that via this resource embedding, a node
$n$ is mapped onto the set $V_n=\{v \in V: \exists r \in R,
\: v=h(key(r)) \wedge (owner(r)=n)\}$, namely the set of
metric-space points assigned to the resources the node provides.

Our next step is to carefully construct a directed random graph
from the points embedded in $V$.
We assume that each newly-arrived node $n$ is initially connected
to some other node in $N$.
Each node $n$ generates the outgoing links for each vertex $v
\in V_n$ independently.
A link $(v,u) \in V_n \times V_m$ simply denotes that $n$ knows that
$m$ is the network node that provides the resource mapped to
$u$; hence, we can view the graph as a virtual overlay network  
of information, pieces of which are stored locally at each node.
Node $n$ constructs each link by executing the search algorithm to locate
the resource that is mapped to the sink of that link. If the metric
space is not populated densely enough, the choice of a sink may
result in a vertex corresponding to an absent resource. In that
case, $n$ chooses the neighbor present closest to the original sink.
Moving to nearby vertices will introduce some bias in the link 
distribution, but the magnitude of error does not appear to be large.
A more detailed description of the graph construction
is given in Section~\ref{sec:RANDOMGRAPHS}.

Having constructed the overlay network of information, we 
can now use it for resource location. As new nodes arrive, 
old nodes depart, and existing ones alter the set of 
resources they provide or even crash, the resources available 
in the distributed database change. At any time $t$, let
$R^t \subseteq R$ be the set of available resources and $I^t$ be
the corresponding overlay network.  A request by node
$n$ to locate resource $r$ at time $t$ is served in a simple,
localized manner: $n$ calculates the metric-space point $v$ that   
corresponds to $r$, and a request message is then routed over  
$I^t$ from the vertex in $V_n$ that is closest to $v$ to $v$   
itself.\footnote{Note that since $R^t$ generally changes with 
time, and may specifically change while the request is being 
served, the request message may be routed over a series of 
different overlay networks $I^{t_1},\:I^{t_2},\: \ldots,\:
I^{t_k}$.}
Each node needs only local information, namely its set
of neighbors in $I^t$, to participate in the resource location.
Routing is done greedily by forwarding the message to the node mapped
to a metric-space point as close to $v$ as possible. The problem of
resource location is thus translated into routing on random graphs
embedded in a metric space.

To a first approximation, our approach is similar to the 
``small-world" routing work by Kleinberg~\cite{KL99}, in which points 
in a two-dimensional grid are connected by links drawn from a normalized
power-law distribution (with exponent 2), and routing is done by having 
each node route a packet to its neighbor closest to the packet's 
destination.
Kleinberg's approach is somewhat brittle because it assumes a 
constant number of links leaving each node. Getting good 
performance using his technique
depends both on having a complete two-dimensional
grid of nodes and on very carefully adjusting the exponent of
the random link distribution. We are not as interested in keeping
the degree down and accept a larger degree to get more 
robustness. We also cannot assume a complete grid: since 
fault-tolerance is one of our main objectives, and since nodes
are mapped to points in the metric space based on what resources
they provide, there may be missing nodes.

The use of random graphs is partly motivated by a desire to keep 
the data structure scalable and the routing algorithm as 
decentralized as possible, as random graphs can be constructed 
locally without global coordination. Another important reason 
is that random graphs are by nature robust against node failures: 
a node-induced subgraph of a random graph is generally still a 
random graph; therefore, the disappearance of a vertex, along with 
all its incident links (due to failure of one of the machines 
implementing the data structure) will still allow routing
while the repair mechanism is trying to heal the damage. 
The repair mechanism also benefits from the use of random graphs, 
since most random structures require less work to maintain their 
much weaker invariants compared to more organized data structures.

Embedding the graph in a metric space has the very important
property that the only information needed to locate a resource
is the location of its corresponding metric-space point. That
location is permanent, both in the sense of being unaffected by
disruption of the data structure, and easily computable by any
node that seeks the resource. So, while the pattern of links
between nodes may be damaged or destroyed by failure of nodes or
of the underlying communication network, the metric space forms
an invulnerable foundation over which to build the ephemeral
parts of the data structure.

\section{Current peer-to-peer systems}
\label{sec:RELATED}

Most of the peer-to-peer systems in widespread use are
not scalable. Napster~\cite{NP} has a central server that services
requests for shared resources even though the actual
resource transfer takes place between the peer requesting
the resource and the peer providing it, without involving
the central authority. However, this has several
disadvantages including a vulnerable single point of
failure, wasted computational power of the clients as
well as not being scalable. Gnutella~\cite{GN} 
floods the network to locate a resource.  Flooding creates a
trade-off between overloading every node in the network
for each request and cutting off searches before completion.
While the use of super-peers \cite{MOR} ameliorates the problem
somewhat in practice, it does not improve performance in the limit.

Some of these first-generation systems
have inspired the development of more sophisticated ones
like CAN~\cite{SR01}, Chord~\cite{CH01} and Tapestry~\cite{TP01}.
CAN partitions a $d$-dimensional metric space into {\em zones}.
Each key is mapped to a point in some zone and stored at the node
that owns the zone.
Each node stores $O(d)$ information, and resource location,
done by greedy routing, takes $O(dn^{1/d})$ time. Chord maps nodes
to identities of $m$ bits placed around a {\em modulo
$2^m$ identifier circle}. Resources are stored at existing
{\em successor} nodes of the nodes they are mapped to. Each node
stores a routing table with $m$ entries such that the $i$-th entry 
stores the key of the first node succeeding it by at least
$2^{i-1}$ on the identifier circle. Each resource is also 
mapped onto the identifier circle and stored at the first
node succeeding the location that it maps to. Routing is 
done greedily to the farthest possible node in the 
routing table, and it is not hard to see that this gives an
$O(\log n)$ delivery time with $n$ nodes in the system.
Tapestry uses Plaxton's algorithm~\cite{PL97}, a form of
suffix-based, hypercube routing, as the routing mechanism:
in this algorithm, the message is forwarded deterministically 
to a node whose identifier is one digit closer to the 
target identifier. To this end, each node maintains $O(\log n)$ 
pieces of information and delivery time is also $O(\log n)$.

Although these systems seem vastly different, there is a recurrent
underlying theme in the use of some variant of an overlay
metric space in which the nodes are embedded. The location
of a resource in this metric space is determined by its key.
Each node maintains some information about its neighbors
in the metric space, and routing is then simply done by
forwarding packets to neighbors {\em closer} to the target
node with respect to the metric.
In CAN, the metric space is explicitly defined 
as the coordinate space which is covered by the zones and 
the distance metric used is simply the Euclidean distance.
In Chord, the nodes can be thought of being 
embedded on grid points on a real circle, with distances measured 
along the circumference of the circle providing the required 
distance metric. In Tapestry, we can think of the nodes being
embedded on a real line and the identifiers are simply the
locations of the nodes on the real line. Euclidean distance
is used as the metric distance for greedy forwarding
to nodes with identifiers closest to the target node.
This inherent common structure leads to similar results
for the performance of such networks. In this paper, we 
explain why most of these systems achieve similar
performance guarantees by 
describing a general setting for such overlay metric spaces,
although most of our results apply only in one-dimensional
spaces.

In general, the fault-tolerance properties of these systems are
not well-defined. Each system provides a repair mechanism for
failures but makes no performance guarantees till this mechanism
kicks in.  For large systems, where nodes appear and
leave frequently, resilience to repeated and concurrent failures
is a desirable and important property. Our experiments show that with
our overlay space and linking strategies, the system performs
reasonably well even with a large number of failures.

\section{Routing}
\label{sec:ROUTING}

In this section, we present our lower and upper bounds on routing.
We consider greedy routing in a graph embedded in a line where  
each node is connected to its immediate neighbors and to multiple
long-distance neighbors chosen according to a fixed link distribution. 
We give lower bounds for greedy routing for \buzz{any} link 
distribution satisfying certain properties 
(Theorem~\ref{theorem-lower-bound}). We also present upper bounds 
in the same model where the long-distance links are chosen as per 
the inverse power-law distribution with exponent $1$ and analyze
the effects on performance in the presence of failures. 

\subsection{Tools}

Some of our upper bounds will be proved using 
a well-known upper bound of Karp~\etal~\cite{KarpUW1988} 
on probabilistic recurrence relations.  
We will restate this bound as
Lemma~\ref{lemma-probabilistic-recurrence-ub}, and then show how
a similar technique can be used to get \emph{lower bounds} with some
additional conditions in Theorem~\ref{theorem-mean-lower-bound}.

\begin{lemma}[\cite{KarpUW1988}]
\label{lemma-probabilistic-recurrence-ub}
The time $T(X_0)$ 
needed for a nonincreasing real-valued Markov chain
$X_0, X_1, X_2, X_3\ldots$ to drop to $1$ is bounded by
\begin{equation}
\label{eq-karp}
T(X_0) \le \int_{1}^{X_0} \frac{1}{\mu_z} dz,
\end{equation}
when $\mu_z = \E[X_t - X_{t+1} : X_t = z]$ is a nondecreasing
function of $z$.
\end{lemma}

This bound has a nice physical interpretation.  If it takes one
second to jump down $\mu_x$ meters from $x$, then we are traveling at a
rate of $\mu_x$ meters per second during that interval.  When we zip
past some position $z$, we are traveling at the average speed $\mu_x$
determined by our starting point $x \ge z$ for the interval.  Since
$\mu$ is nondecreasing, using $\mu_z$ as our estimated speed
underestimates our actual speed when passing $z$.  The integral
computes the time to get all the way to zero if we use
$\mu_z$ as our instantaneous speed when passing position $z$.  Since
our estimate of our speed is low (on average), our estimate of our time
will be high, giving an upper bound on the actual expected time.

We would like to get lower bounds on such processes in
addition to upper bounds, and we will not necessarily be able to
guarantee that $\mu_z$, as defined in
Lemma~\ref{lemma-probabilistic-recurrence-ub}, will be a
nondecreasing function of $z$.  But we will still use the same basic
intuition: The average speed at which we pass $z$ is at most the
maximum average speed of any jump that takes us past $z$.  We can find
this maximum speed by taking the maximum over all $x > z$;
unfortunately, this may give us too large an estimate.  Instead, we
choose a threshold $U$ for ``short'' jumps,
compute the maximum speed of short jumps of at most $U$ for 
all $x$ between $z$ and $z+U$, and
handle the (hopefully rare) long jumps of more than $U$ by
conditioning against them.  Subject to this conditioning, we can
define an upper bound $m_z$ on the average speed passing $z$, and
use essentially the same integral as in (\ref{eq-karp}) to get a lower
bound on the time.  Some additional tinkering to account for the
effect of the conditioning then gives us our real lower bound,
which appears in Theorem~\ref{theorem-mean-lower-bound} below, as 
Inequality (\ref{eq-mean-lower-bound}).

\newcommand{\dft}{f(X_t) - f(X_{t+1})}
\newcommand{\dyt}{Y_t - Y_{t+1}}
\newcommand{\dzt}{Z_t - Z_{t+1}}
\newcommand{\ftat}{\calF_t, A_t}
\newcommand{\ef}[1]{\E\left[{#1}:\ftat\right]}
\newcommand{\muf}{\mu_{f(X_t)}}
\newcommand{\ydenom}{\epsilon Y_0 + (1-\epsilon)}
\newcommand{\yydenom}{\left(\epsilon Y_0 + (1-\epsilon)\right)}

\begin{theorem}
\label{theorem-mean-lower-bound}
Let $X_0, X_1, X_2, \ldots$ be 
Markov process with state space $S$, where
$X_0$ is a constant.
Let $f$ be a non-negative real-valued function on $S$
such that, for all $t$,
\begin{equation}
\label{eq-nonincreasing}
\Pr[\dft \ge 0 : X_t] = 1.
\end{equation}
Let $U$ and $\epsilon$ be constants such that for any $x > 0$,
\begin{equation}
\label{eq-mean-lower-bound-U-epsilon}
\Pr[\dft \ge U : X_t = x] \le \epsilon.
\end{equation}
Let 
\begin{equation}
\label{eq-mean-lower-bound-tau}
\tau = \min \{ t: f(X_t) = 0 \}.
\end{equation}
For each $x$ with $f(x) > 0$, let $\mu_x > 0$ satisfy
\begin{equation}
\label{eq-mean-lower-bound-mu}
\mu_x \ge \E[\dft : X_t = x, \dft < U].
\end{equation}
Now define
\begin{equation}
\label{eq-mean-lower-bound-m}
m_z = \sup \left\{ \mu_x: x\in S, f(x) \in [z, z+U) \right\},
\end{equation}
and define
\begin{equation}
\label{eq-mean-lower-bound-T}
T(x) = \int_{0}^{f(x)} \frac{1}{m_z} dz.
\end{equation}
Then
\begin{equation}
\label{eq-mean-lower-bound}
\E[\tau] \ge \frac{T(X_0)}{\epsilon T(X_0) + (1-\epsilon)}.
\end{equation}
\end{theorem}
\begin{proof}
Define
\begin{equation}
\label{eq-mean-lower-bound-y}
Y_t = \left\{
\begin{array}{cl}
T(X_t) & \mbox{, if $f(X_{t'}) - f(X_{t'+1}) < U$ for all $t' < t$,} \\
0      & \mbox{, otherwise.}
\end{array}
\right.
\end{equation}
The idea is that $Y_t$ drops to zero immediately if a long jump
occurs.  We will show that even with such overeager jumping, $Y_t$
does not drop too quickly on average.  The intuition is that the chance of a 
long jump reduces
$Y_t$ by at most an expected $\epsilon Y_t \le \epsilon Y_0$, while
the effect of short jumps can be bounded by applying the definition of
$T$.

Let $\calF_t$ be the $\sigma$-algebra
generated by $X_0, X_1, \ldots X_t$.  Let $A_t$ be the event that
$\dft < U$, that is, that the jump from $f(X_t)$ to $f(X_{t+1})$ is a
short jump.
Now compute
\begin{eqnarray}
E\left[\dyt:\calF_t\right]
&=&
\Pr\left[\,\overline{A_t}:\calF_t\right] (Y_t - 0)
+ (1 - \Pr\left[\,\overline{A_t}:\calF_t\right]) \ef{\dyt}\nonumber
\\
&\le&
\Pr\left[\,\overline{A_t}:\calF_t\right] Y_0
+ (\epsilon - \Pr\left[\,\overline{A_t}:\calF_t\right]) Y_0
+ (1-\epsilon) \ef{\dyt}\nonumber
\\
&=&
\epsilon Y_0 + (1-\epsilon) \ef{\dyt}.\label{eq-mean-lower-bound-dyt}
\end{eqnarray}

Now let us bound $\ef{\dyt}$.  Expanding the definitions
(\ref{eq-mean-lower-bound-T}) and (\ref{eq-mean-lower-bound-y})
gives
\begin{equation}
\label{eq-mean-lower-bound-integral-expansion}
\ef{\dyt}
=
\ef{\int_{f(X_{t+1})}^{f(X_t)} \frac{1}{m_z} dz}.
\end{equation}

Now, conditioning on $A_t$ means that 
$f_(X_{t+1}) > f(X_t) - U$
and thus
$z > f(X_t) - U$ for the entire range of the integral. 
It follows that $f(X_t)$ lies in the half-open interval $[z,z+U)$ for
each such $z$, from which we have $m_z \ge \muf$
from (\ref{eq-mean-lower-bound-m}).
Inverting gives $\frac{1}{m_z} \le \frac{1}{\muf}$,
and plugging this inequality into
(\ref{eq-mean-lower-bound-integral-expansion}) gives
\begin{eqnarray}
\ef{\dyt}
&\le&
\ef{\int_{f(X_{t+1})}^{f(X_t)} \frac{1}{\muf} dz}
\nonumber
\\
&=&
\frac{1}{\muf} \ef{\dft}
\nonumber
\\
&\le&
\frac{1}{\muf} \muf
\nonumber
\\
&=&
1.
\label{eq-mean-lower-bound-dytat}
\end{eqnarray}

Applying (\ref{eq-mean-lower-bound-dytat})
to
(\ref{eq-mean-lower-bound-dyt}) gives
\begin{equation}
\label{eq-mean-lower-bound-dyt-final}
\E[\dyt : \calF_t ]
\le
\ydenom.
\end{equation}

We have now shown that $Y_t$ drops slowly on average.  To turn this
into a lower bound on the time at which it first reaches zero, define
$Z_t = Y_t + \min(t, \tau) \yydenom$.
Conditioning on $t < \tau$, observe that
\begin{eqnarray*}
\E[\dzt:\calF_t, t < \tau]
&=&
\E[\dyt:\calF_t, t < \tau] - \yydenom
\\
&\le&
\yydenom - \yydenom 
\\
&=&
0.
\end{eqnarray*}

Alternatively, if $t \ge \tau$ we have
\begin{displaymath}
\E[\dzt:\calF_t, t \ge \tau] = 0.
\end{displaymath}

In either case, $\E[\dzt:\calF_t] \le 0$, implying
$Z_t \le \E[Z_{t+1}:\calF_t]$.
In other words, $\{Z_t, \calF_t\}$ is a submartingale.

Because $\{Z_t, \calF_t\}$ is a submartingale, and $\tau$ is a
stopping time relative to $\{\calF_t\}$, we have
$Z_0 = Y_0 
\le \E[Z_\tau] 
= \E\left[ 0 + \tau \yydenom \right]
= \yydenom \E[\tau]$.
Solving for $\E[\tau]$ then gives
\begin{displaymath}
\E[\tau] \ge \frac{Y_0}{\ydenom} 
= \frac{T(X_0)}{\epsilon T(X_0) + (1-\epsilon)}.
\end{displaymath}
\end{proof}

\subsection{Lower bounds on greedy routing}

We will now show a lower bound on the expected time taken by greedy
routing on a random graph embedded in a line. Each node in the
graph has expected outdegree at most $\ell$ and is connected to its 
immediate neighbor on either side. For polylogarithmic values of $\ell$,
we consider two variants of the greedy routing algorithm and derive lower 
bounds for them equal to $\Omega(\log^2 n / (\ell^2 \log \log n))$ and to
$\Omega(\log^2 n / (\ell \log \log n))$, as stated in 
Theorem~\ref{theorem-lower-bound}.
The routing variants, along with the machinery and proofs of the
associated lower bounds, are presented in Sections~\ref{Section-lower-bound} 
through \ref{Section-putting-the-pieces-together}. For large values
of $\ell$, a lower bound of $\Omega(\frac{\lg n}{\lg \ell})$
on the worst-case routing time can be 
derived very simply, as follows.

\begin{theorem}
\label{theorem-tree-lower-bound}
Let $\ell \in (\lg n, n^c]$. Then for any link distribution and any
routing strategy, the
delivery time $T = \Omega(\frac{\log n}{\log \ell})$.
\end{theorem}
\begin{proof}
With $\ell$ links for each node, we can reach at 
most $\ell^k$ nodes at step $k$. Assuming that the minimum time to 
reach all $n$ nodes is T, $\ell^T = n$. This gives a lower bound of 
$\Omega(\frac{\log n}{\log \ell})$ on $T$.
\end{proof}

\subsubsection{Lower bound for polylogarithmic number of links}
\label{Section-lower-bound}

We consider the case of the expected outdegree of each node falling in
the range $[1,\lg n]$. The probability that a node at
position $x$ is connected to positions $x-\Delta_1, x-\Delta_2,
\ldots, x-\Delta_k$ depends only on the set $\Delta=\{\Delta_1, \ldots,
\Delta_k\}$ and not on $x$ and is independent of the choice of outgoing
links for other nodes.\footnote{We assume that nodes are labeled by
integers and identify each node with its label to avoid excessive
notation.} Since we assume that each node is connected to its immediate
neighbors, we require that $\pm 1$ appears in $\Delta$. 

We consider two variants of the greedy routing algorithm. Without
loss of generality, we assume that the target of the search is labeled
$0$. In \buzz{one-sided greedy routing}, the algorithm never traverses a 
link that would take it past its target.  So if the algorithm is currently
at $x$ and is trying to reach $0$, it will move to the node $x-\Delta_i$ 
with the smallest non-negative label.  In \buzz{two-sided greedy routing}, 
the algorithm chooses a link that minimizes the distance to the target, 
without regard to which side of the target the other end of the link is.  
In the two-sided case the algorithm will move to a node $x-\Delta_i$ 
whose label has the smallest absolute value, with ties broken 
arbitrarily. One-sided greedy routing can be thought of as modeling 
algorithms on a graph with a boundary when the target lies on the 
boundary, or algorithms where all links point in only one direction 
(as in Chord).

Our results are stronger for the one-sided case than for the two-sided
case.  With one-sided greedy routing, we show a lower bound of
$\Omega(\log^2 n / (\ell \log \log n))$ on the time to reach $0$ from a 
point chosen uniformly from the range $1$ to $n$ that applies to any link
distribution.  For two-sided routing, we show a lower bound of
$\Omega(\log^2 n / (\ell^2 \log \log n))$, with some constraints on the 
distribution.  We conjecture that these constraints are unnecessary, and 
that $\Omega(\log^2 n / (\ell \log \log n))$ is the correct lower bound 
for both models. A formal statement of these results appears as 
Theorem~\ref{theorem-lower-bound} in 
Section~\ref{Section-putting-the-pieces-together}, but before we can 
prove it we must develop machinery that will be useful in the
proofs of both the one-sided and two-sided lower bounds.

\subsubsection{Link sets: notation and distributions}

First we describe some notation for $\Delta$ sets.
Write each $\Delta$ as 
\[\{\Delta_{-s}, \ldots \Delta_{-2}, \Delta_{-1} = -1, 
 \Delta_{1} = 1, \Delta_{2}, \ldots  \Delta_{t}\},\]
where $\Delta_{i} < \Delta_{j}$ whenever $i < j$.
Each $\Delta$ is a random variable drawn from some distribution on
finite sets; the individual $\Delta_i$ are thus in general \emph{not}
independent.
Let $\Delta^-$ consist of the $s$ negative elements of $\Delta$
and $\Delta^+$ consist of the $t$ positive elements.
Formally define $\Delta_{-i} = -\infty$ when $i > s$ 
and $\Delta_{i} = +\infty$ when $i > t$.

For one-sided routing, we make no assumptions about the distribution
of $\Delta$ except that $|\Delta|$ must have finite expectation and
$\Delta$ always contains $1$.  For two-sided routing, we assume that
$\Delta$ is generated by including each possible $\delta$ in $\Delta$
with probability $p_\delta$, where $p$ is symmetric about the origin
(i.e., $p_\delta = p_{-\delta}$ for all $\delta$),
$p_1 = p_{-1} = 1$, and $p$ is
unimodal, i.e. nonincreasing for positive $\delta$ and nondecreasing
for negative $\delta$.\footnote{These constraints imply
that $p_0 = 1$;
formally, we imagine that $0$ is present in each $\Delta$ but is
ignored by the routing algorithm.}  We also require that the events
$[\delta \in \Delta]$ 
and
$[\delta' \in \Delta]$
are pairwise independent for distinct $\delta,\delta'$.

\subsubsection{The aggregate chain $S^t$}

For a fixed distribution on $\Delta$, the trajectory
of a single initial point $X^0$ is a Markov chain $X^0, X^1, X^2, \ldots$,
with $X^{t+1} = s(X^t, \Delta^t)$,
where $\Delta^t$ determines the outgoing links from the node reached
at time $t$ and $s$ is a \buzz{successor function} that selects the next node
$X^{t+1} = X^t - \Delta^t_i$
according to the routing algorithm.
Note that the chain is Markov, because the presence of $\pm 1$ links
guarantees that no node ever appears twice in the sequence, and so
each new node corresponds to a new choice of links.

\newcommand{\Di}{{\Delta i}}
\newcommand{\Dis}{{\Delta i \sigma}}

From the $X^t$ chain we can derive an \buzz{aggregate chain}
that describes the
collective behavior of all nodes in some
range.  
Each state of the aggregate chain is a contiguous sets of nodes whose
labels all have the same sign;
we define the sign of the state to be the common sign of all of its
elements.
For one-sided routing each state is either $\{0\}$ or an interval
of the form $\{1\ldots k\}$ for some $k$.  For two-sided routing the
states are more general
The aggregate states are characterized formally in
Lemma~\ref{lemma-aggregate-ranges}.

Given a contiguous set of nodes $S$ and a set $\Delta$,
define
\begin{displaymath}
S_{\Di} = \{ x \in S : s(x, \Delta) = x - \Delta_i \}.
\end{displaymath}
The intuition is that $S_{\Di}$ consists of all those nodes for which
the algorithm will choose $\Delta_i$ as the outgoing link.
Note that $S_{\Di}$ will always be a contiguous range because of the
greediness of the algorithm.
Now define, for each $\sigma \in \{-, 0, +\}$:
\begin{displaymath}
S_{\Dis} = \{ x \in S_{\Di} : \sgn s(x, \Delta) = \sigma \}.
\end{displaymath}
Here we have simply split $S_{\Di}$ into those nodes with
negative, zero, or positive successors.

For any set $A$ and integer $\delta$ write $A-\delta$
for $\{x-\delta : x \in A\}$.

We will now build our aggregate chain by letting 
the successors of a range $S$ be the ranges $S_{\Dis}-\Delta_i$ 
for all possible
$\Delta$, $i$, and $\sigma$.  
As a special case, we define $S^{t+1} = \{0\}$ when $S^{t} = \{0\}$;
once we arrive at the target, we do not leave it.
For all other $S^t$, we let
\begin{equation}
\label{eq-stdis-prob}
\Pr\left[S^{t+1} = S^t_{\Dis} - \Delta_i : \Delta\right] 
= \frac{|S^t_{\Dis}|}{|S^t|},
\end{equation}
and define the unconditional transition probabilities by averaging
over all $\Delta$.

Lemma~\ref{lemma-aggregate-chain-works} shows that moving to the
aggregate chain does not misrepresent the underlying single-point
chain:

\begin{lemma}
\label{lemma-aggregate-chain-works}
Let $X^0$ be drawn uniformly from the range $S^0$.  Let $Y^t$ be a
uniformly chosen element of $S^t$.  Then for all $x$ and $t$,
$\Pr[X^t = x] = \Pr[Y^t = x]$.
\end{lemma}
\begin{proof}
Clearly the lemma holds for $t=0$.
Fix $S^{t-1}$, and consider two methods for generating $Y^{t}$.  
The first generates $Y^t$ directly from $Y^{t-1}$ and
shows that $Y^t$ generated in this way has the same distribution as
$X^t$.
The second generates $Y^t$ from $S^t$ as describe in the lemma
and produces the same
distribution on $Y^t$ as the first.

In the first method, 
we choose $Y^{t-1}$ uniformly from $S^{t-1}$, choose a
random $\Delta^{t-1}$, and compute $s(Y^{t-1}, \Delta^{t-1}$.
Here the transition rule applied to $Y^{t-1}$ is the same as for
$X^{t-1}$, so under the induction hypothesis that $Y^{t-1}$ and
$X^{t-1}$ are equal in distribution, so are $Y^t$ and $X^t$.

In the second method, we again choose a random $\Delta^{t-1}$ 
and then choose $S^{t}$ by choosing some $S^{t-1}_{\Dis}$ in proportion
to its size, let $S^{t} = S^{t-1}_\Dis - \Delta_i$, and then let $Y^t$
be a uniformly chosen element of $S^t$.
We can implement the choice of $S^{t-1}_\Dis$ by choosing some $Y^{t-1}$
uniformly from $S^{t-1}$ and picking $S^{t-1}_\Dis$ as the subrange
that contains $Y^{t-1}$; and we can simplify the task of choosing
$Y^{t}$ by setting it equal to $Y^{t-1} - \Delta_i$, since
conditioning on $Y^{t-1} \in S^{t-1}_\Dis$ leaves $Y^{t-1}$ with a
uniform distribution.  But by implementing the second method in this
way, we have reduced it to the first, and the lemma is proved.
\end{proof}

Lemma~\ref{lemma-aggregate-ranges} justifies our earlier
characterization of the aggregate state spaces:

\begin{lemma}
\label{lemma-aggregate-ranges}
Let $S^0 = \{ 1 \ldots n \}$ for some $n$.
Then with one-sided routing,
every $S^t$ is either $\{0\}$ or of the form $\{1\ldots k\}$ for some
$k$; 
and with two-sided routing,
every $S^t$ is an interval of integers in which every element has the
same sign.
\end{lemma}
\begin{proof}
By induction on $t$.  For one-sided routing, observe that
$S^{t-1}_{\Di -}$ is always empty, as the routing algorithm is not
allowed to jump to negative nodes.  If $S^t = S^{t-1}_{\Di 0} -
\Delta_i$, then 
$S^t = \{\Delta_i\} - \Delta_i = \{0\}$.
Otherwise $S^t = S^{t-1}_{\Di +} - \Delta_i$; but since 
$S^{t-1} = \{1 \ldots k\}$ for some $k$,
if it contains any point $x$ greater than $\Delta_i$ it must contain
$\Delta_i + 1$; thus $\min(S^{t-1}_{\Di +} = \Delta_i + 1$
and so $\min(S^t)$ becomes $1$.

The result for the two-sided case is immediate from the fact that
$S^{t} = S^{t-1}_\Dis - \Delta_i$
combined with the definition of $S^{t-1}_\Dis$.
\end{proof}

The advantage of the aggregate chain over the single-point chain is
that, while we cannot do much to bound the progress of a single point
with an arbitrary distribution on $\Delta$, we can show that the size
of $S^t$ does not drop too quickly given a bound $\ell$ on
$\E[|\Delta|]$.
The intuition is that each successor
set of size $a^{-1} |S^t|$ or less occurs
with probability at most $a^{-1}$, and there are at most $3\ell$ such
sets on average.

\newcommand{\Prsta}{\Pr\left[|S^{t+1}| \le a^{-1} |S^t| : S^t\right]}
\begin{lemma}
\label{lemma-aggregate-max-drop}
Let $\E[|\Delta|] \le \ell$.  Then for any $a \ge 1$,
in either the one-sided or two-sided model,
\begin{equation}
\label{eq-aggregate-max-drop}
\Prsta \le 3\ell a^{-1}.
\end{equation}
\end{lemma}
\begin{proof}
\begin{sloppypar}
Fix $S^t$.
First note that if $a^{-1} |S^t| < 1$, then $\Prsta = 0$.
So we can assume that $a^{-1} |S^t| \ge 1$ and in particular that
$a \le |S^t|$.
\end{sloppypar}

Conditioning on $\Delta$, there are at most $3|\Delta|$ non-empty sets
$S^t_{\Dis}$.  
If $|S^t_\Dis| \le a^{-1} |S^t|$, then $|S^t_\Dis|$ is chosen with
probability at most $a^{-1}$ by (\ref{eq-stdis-prob}).
Thus the probability of choosing any of the at most $3|\Delta|$ sets
$S^t_\Dis$ of size at most $a^{-1}|S^t|$ is at most $3|\Delta|a^{-1}$.

Now observe that
\begin{eqnarray*}
\Prsta &\le&
    \sum_d
        \Pr\left[ |\Delta| = d \right] 3da^{-1} \\
    &=& 3a^{-1} \E\left[|\Delta|\right] \\
    &\le& 3 \ell a^{-1}.
\end{eqnarray*}
\end{proof}

\begin{sloppypar}
Another way to write (\ref{eq-aggregate-max-drop}) is to say that
$\Pr\left[ \ln |S^t| - \ln |S^{t+1}| \ge \ln a : S^t \right] \le 3 \ell
a^{-1}$, which will give the bound 
(\ref{eq-mean-lower-bound-U-epsilon}) on the probability of large
jumps when it comes time to apply
Theorem~\ref{theorem-mean-lower-bound}.
\end{sloppypar}

\subsubsection{Boundary points}
\label{section-boundary-points}

Lemma~\ref{lemma-aggregate-max-drop} says that $|S^t|$ seldom drops by
too large a ratio at once, but it doesn't tell us much about how
quickly $|S^t|$ drops in short hops.  To bound this latter quantity,
we need to get a bound on how many subranges $S^t$ splinters into
through the action of $s(\cdot, \Delta)$.
We will do so by showing that only certain points can appear as the
boundaries of these subranges in the direction of $0$.

For fixed $\Delta$, define for each $i > 0$
\begin{displaymath}
\beta_i = \ceil{\frac{\Delta_i+\Delta_{i+1}}{2}}
\end{displaymath}
and
\begin{displaymath}
\beta_{-i} = \floor{\frac{\Delta_{-i}+\Delta_{-i-1}}{2}}.
\end{displaymath}
Let $\beta$ be the set of all finite $\beta_i$ and $\beta_{-i}$.

\begin{lemma}
\label{lemma-boundary-points}
Fix $S$ and $\Delta$ and let $\beta$ be defined as above.
Suppose that $S$ is positive.
Let $M = \{ \min(S_\Dis) : S_\Dis \ne \emptyset \}$ be the set of
minimum elements of subranges $S_\Dis$ of $S$.
Then $M$ is a subset of $S$ and contains no elements other than
\begin{enumerate}
\item $\min(S)$,
\item $\Delta_i$ for each $i > 0$, 
\item $\Delta_i+1$ for each $i > 0$, and
\item at most one of $\beta_i$ or $\beta_i+1$ for each $i > 0$,
\end{enumerate}
where the last case holds only with two-sided routing.

If $S$ is negative, the symmetric condition holds for
$M = \{ \max(S_\Dis) : S_\Dis \ne \emptyset \}$.
\end{lemma}
\begin{proof}
Consider some subrange $S_\Dis$ of $S$.  If $S_\Dis$ contains
$\min(S)$, the first case holds.  Otherwise:
(a) if $S_\Dis = S_{\Di 0}$, the second
case holds; (b) if $S_\Dis = S_{\Di +}$, the third case holds;
(c) if $S_\Dis = S_{\Di -}$, the fourth case holds, with $\min(S_{\Di
-}) = \beta_{i-1}$ if $\Delta_{i-1} + \Delta_i$ is odd, and either
$\beta_{i-1}$ or $\beta_{i-1}+1$ if $\Delta_{i-1} + \Delta_i$ is even,
depending on whether the tie-breaking rule assigns $\beta_{i-1}$ to
$S_{\Delta(i-1)+}$ or $S_{\Di -}$.
\end{proof}

We will call the elements of $M$ \buzz{boundary points} of $S$.

\subsubsection{Bounding changes in $\ln |S^t|$}

Now we would like to use Lemmas~\ref{lemma-aggregate-max-drop} and
Lemma~\ref{lemma-boundary-points} to get an upper bound on the rate at
which $\ln |S^t|$ drops as a function of the $\Delta$ distribution.

The following lemma is used to bound a sum that arises in
Lemma~\ref{lemma-log-drop}.

\begin{lemma}
\label{lemma-conditional-convex}
Let $c \ge 0$.
Let $\sum_{i=1}^{n} x_i = M$ where each $x_i \ge 0$ and at least one
$x_i$ is greater than $c$
Let $B$ be the set of all $i$ for which $x_i$ is greater than $c$.
Then
\begin{equation}
\label{eq-condition-convex}
\frac{
  \sum_{i \in B} x_i \ln x_i
}{
  \sum_{i \in B} x_i
}
\ge
\ln\left( \max \left(c, \frac{M}{n}\right)\right).
\end{equation}
\end{lemma}
\begin{proof}
If $\frac{M}{n} < c$, 
we still have $x_i > c$ for all
$i \in B$, so the left-hand side cannot be
less than $\ln c$.
So the interesting
case is when $\frac{M}{n} > c$.

Let $B$ have $b$ elements.  Then $\sum_{i \notin B} x_i < (n-b)c$
and $\sum_{i \in B} \ge M - (n-b)c = M-nc+bc$.
Because $x_i \ln x_i$ is convex, its sum over $B$ is minimized for fixed
$\sum_{i\in B} x_i$ by setting all such $x_i$ equal, in which case the
left-hand side of (\ref{eq-condition-convex}) becomes simply
$\ln(x_i)$ for any $i \in B$.

Now observe that setting all $x_i$ in $B$ equal gives
$x_i = \frac{M-nc+bc}{b} 
= \frac{M-nc}{b} + c 
\ge \frac{M-nc}{n} + c
= \frac{M}{n}$.
\end{proof}

\newcommand{\aS}{a^{-1}|S|}
\newcommand{\lndrop}{\ln|S^t|-\ln|S^{t+1}|}
\newcommand{\constdrop}{\ln\frac{1}{1 - a^{-1}}}
\begin{lemma}
\label{lemma-log-drop}
Fix $a > 1$, and 
let $S = S^{t}$ be a positive range with $|S| \ge a$.
Define $\beta$ as in Lemma~\ref{lemma-boundary-points}.
Let $S' = [\min(S) + \ceil{\aS} - 1, \max(S)-1]$.
Let $A$ be the event $\left[\lndrop < \ln a\right]$.
Then
\begin{equation}
\label{eq-log-drop}
\E \left[ \lndrop : S^t, A \right]
\le
\constdrop + \frac{\ln \E[1+Z : S^t]}{\Pr[A: S^t]},
\end{equation}
where $Z = 2|\Delta \cap S'|$ with one-sided routing
and $Z=2|\Delta \cap S'| + |\beta \cap S'|$ with two-sided routing.
\end{lemma}
\begin{proof}
Call a subrange $S_\Dis$ \buzz{large} if $|S_\Dis| > \aS$ and
\buzz{small} otherwise; the intent is that the large ranges are
precisely those that yield $\lndrop < \ln a$.
Observe that for any large $S_\Dis$, $|S_\Dis| > \aS \ge 1$, 
implying any large set has at least two elements.

For any large $S_\Dis$, 
$\max(S_\Dis) 
 \ge \min(S) + \ceil{\aS} - 1$.
Similarly
$\min(S_\Dis)
 \le \max(S) - 1$.
So any large $S_\Dis$ intersects $S'$ in at least one point.

Let $T = \{T_1, T_2, \ldots, T_k\}$ 
be the set of subranges $S_\Dis$, large or small, that
intersect $S'$.  It is immediate from this definition
that $\bigcup T \supseteq S'$ and thus $\sum |T_j| \ge |S'|$.

Using Lemma~\ref{lemma-boundary-points}, we can characterize the
elements of $T$ as follows.
\begin{enumerate}
\item There is at most one set $T_j$ that contains $\min(T_j)$.
\item There is at most one set $T_j$ that has $\min(T_j) = \Delta_i$ for each
$\Delta_i$ in $S'$.
\item There is at most one set $T_j$ that has $\min(T_j) = \Delta_i+1$ for
each $\Delta_i$ in $S'$.
\item With two-sided routing, 
there is at most one set $T_j$ that has $\min(T_j) = \beta_i$ or
$\min(T_j) = \beta_i+1$ for each $\beta_i$ in $S'$.  Note that there
may be a set whose minimum element is $\beta_i+1$ where $\beta_i =
\min(S') - 1$, but this set is already accounted for by the first
case.
\end{enumerate}

Thus $T$ has at most $1+Z = 1+2|\Delta \cap S'|$ elements with one-sided
routing and at most $1+Z = 1+2|\Delta \cap S'| + |\beta \cap S'|$ elements
with two-sided routing.

Conditioning on $|S^{t+1}| > \aS$, 
$|S^{t+1}|$ is equal to $|S_\Dis|$ for some large $S_\Dis$ and thus
for some large $T_j \in T$.
Which large $T_j$ is chosen is proportional to its size, so
for fixed $T$, we have
\begin{eqnarray*}
\E[\ln S^{t+1} : T, A] &=&
\frac{
  \sum_{j=1}^{|T|} |T_j| \ln |T_j|
}{
  \sum_{j=1}^{|T|} |T_j|
} \\
&\ge& \ln\left(\max\left(\aS, \frac{|\bigcup T|}{|T|}\right)\right) \\
&\ge& \ln\left(\frac{|S'|}{|T|}\right),
\end{eqnarray*}
where the first inequality follows from
Lemma~\ref{lemma-conditional-convex}.

Now let us compute
\begin{eqnarray*}
\E[\lndrop : S^t, A ]
&=& \ln|S^t| - \E[\ln|S^{t+1}| : S^t, A] \\
&\le& \ln|S^t| - \E[\ln |S'| - \ln |T| : S^t, A] \\
&=& \ln \frac{|S^t|}{|S'|} + \E[\ln |T| : S^t, A] \\
&\le& \ln \frac{|S^t|}{|S'|} + \frac{\E[\ln |T| : S^t]}{\Pr[A: S^t]} \\
&\le& \constdrop + \frac{\ln \E[|T| : S^t]}{\Pr[A: S^t]}.
\end{eqnarray*}
In the second-to-last step, we use
$\E[\ln |T| : S^t, A] \le \E[\ln |T| : S^t] / \Pr[A: S^t]$,
which follows from
$\E[\ln |T| : S^t] 
= 
 \E[\ln |T| : S^t, A] \Pr[A: S^t]
+\E[\ln |T| : S^t, \neg A] \Pr[\neg A: S^t]$.
In the last step, we use $\E[\ln |T| : S^t, A] \le \ln E[|T| : S^t, A]$,
which follows from the concavity of $\ln$ and Jensen's inequality.
\end{proof}

\subsubsection{Putting the pieces together}
\label{Section-putting-the-pieces-together}

We now have all the tools we need to prove our lower bound.

\newcommand{\ZZ}{\mathbf{Z}} 
\newcommand{\bh}{\hat{\beta}}
\newloglike{\roundfromzero}{absceil}
\newcommand{\rfz}[1]{\roundfromzero\left({#1}\right)}
\begin{theorem}
\label{theorem-lower-bound}
Let $G$ be a random graph whose nodes are labeled by the integers.
Let $\Delta_x$ for each $x$ be a set of integer offsets chosen
independently from some common distribution, subject to the constraint
that $-1$ and $+1$ are present in every $\Delta_x$,
and let node $x$ have an outgoing link to $x-\delta$ for each
$\delta\in\Delta_x$.  Let $\ell = \E[|\Delta|]$.
Consider a greedy routing trajectory in $G$ starting at a point chosen
uniformly from $1 \ldots n$ and ending at $0$.

With one-sided routing, the expected time to reach $0$ is
\begin{equation}
\label{eq-lower-bound-one-sided}
\Omega\left(
      \frac{\log^2 n}{\ell \log \log n}
\right).
\end{equation}

With two-sided routing, the expected time to reach $0$ is
\begin{equation}
\label{eq-lower-bound-two-sided}
\Omega\left(
      \frac{\log^2 n}{\ell^2 \log \log n}
\right),
\end{equation}
provided $\Delta$ is generated by including each
$\delta$ in $\Delta$ with probability $p_\delta$, where (a) $p$ is
unimodal, (b) $p$ is symmetric about $0$, and (c) the choices to
include particular $\delta, \delta'$ are pairwise independent.
\end{theorem}
\begin{proof}
Let $S^0 = \{ 1 \ldots n \}$.

We are going to apply Theorem~\ref{theorem-mean-lower-bound} to the
sequence $S^0, S^1, S^2, \ldots$ with $f(S) = \ln |S|$.
We have chosen $f$ so that when we reach the target, $f(S)=0$; so that
a lower bound on $\tau$ gives a lower bound on the expected time of
the routing algorithm.
To apply the theorem,
we need to
show that (a) the probability that $\ln |S|$ drops by a large amount
is small, and (b) that the integral in
(\ref{eq-mean-lower-bound-T}) is large.

\begin{sloppypar}
Let $a = 3 \ell \ln^3 n$.
By Lemma~\ref{lemma-aggregate-max-drop},
for all $t$,
$\Prsta \le 3 \ell a^{-1} = \ln^{-3} n$,
and thus
$\Pr[\lndrop \ge \ln a : S^t] \le \ln^{-3} n$.
This satisfies (\ref{eq-mean-lower-bound-U-epsilon})
with $U = \ln a$ and $\epsilon = \ln^{-3} n$.
\end{sloppypar}

For the second step,
Theorem~\ref{theorem-mean-lower-bound} requires that we bound the
speed of the change in $f(S)$ solely as a function of $f(S)$.  For
one-sided routing this is not a problem, as
Lemma~\ref{lemma-aggregate-ranges} shows that $f(S)$, which reveals
$|S|$, characterizes $S$ exactly except when $|S| = 1$ and the lower
bound argument is done.  For two-sided routing, the situation is more
complicated; there may be some $S^t$ which is not of the form
$\{1\ldots |S^t|\}$ or $\{0\}$, and we need a bound on the speed at
which $\ln |S^t|$ drops that applies equally to all sets of the same
size.

\begin{sloppypar}
It is for this purpose (and only for this purpose)
that we use our conditions on $\Delta$ for
two-sided routing.
Suppose that each $\delta$
appears in $\Delta$ with probability $p_\delta$, that these
probabilities are pairwise-independent, and that the sequence $p$ is
symmetric and unimodal.
Let $\bh = \left\{ \rfz{\frac{x+y}{2}} : x, y \in \Delta, x \ne y \right\}$,
where $\rfz{z}$, the \buzz{absolute ceiling} of $z$,
is $\ceil{z}$ when $z \ge 0$ and $\floor{z}$ when $z \le 0$.
Observe that $\bh \supseteq \beta$; in effect, we are counting in
$\bh$ all
midpoints of pairs of distinct elements of $\delta$ without regard to 
whether the elements are adjacent.
For each $k$, the expected number of distinct 
pairs $x$, $y$ with $x+y=z$ and
$x,y \in \Delta$ is at most
$b_k = \sum_{i=-\infty}^{\infty} p_{k-i} p_i$;
this is a convolution of the non-negative, symmetric, and unimodal 
$p$ sequence with itself and so it is also symmetric and unimodal.  
It follows that for all $0 \le k < k'$, $b_k \ge b_{k'}$, and similarly
$b_{-k} \ge b_{-k'}$.
\end{sloppypar}

Now for the punch line: for each $\delta \ne 0$, 
$q_\delta = b_{2\delta - \sgn \delta} + b_{2\delta}$
is an upper bound on the expected number of distinct pairs $x,y$ that
put $\delta$ in $\beta$, which is in turn an upper bound on 
$\Pr[\delta \in \beta]$, and from the unimodularity of $b$ we have
that $q_\delta \ge q_{\delta'}$ and $q_{-\delta} \ge q_{-\delta'}$
whenever $0 < \delta < \delta'$.  Though $q$ grossly over counts
the elements of $\beta$ (in particular, it gives a bound on $\E[|\beta|]$
of $\ell^2$), its ordering property means that we can bound the
expected number of elements of $\beta$ that appear in some subrange 
of any positive $S^t$
by using $q$ to bound the expected number of elements that
appear in the corresponding subrange of $\{ 1 \ldots |S^t| \}$, and
similarly for negative $S^t$ and $\{-1 \ldots - |S^t| \}$.
Because $p_i$ already satisfies a similar ordering property, we
can thus bound the number of elements of both $\Delta$ and $\beta$
that hit a fixed subrange of $S^t$ given only $|S^t|$.  We do this next.

For convenience, formally define $p_i = \Pr[i \in \Delta]$ and 
$q_i=0$ for one-sided routing.
We will simplify some of the summations by first summing the $p_i$ and $q_i$
over certain pre-defined intervals.
For each integer $i > 0$ let 
$A_i = \{ k \in \ZZ : a^i-1 \le k < a^{i+1}-1\}
 = \{k \in \ZZ : \floor{\ln_a k+1} = i \}$.
Let $\gamma_i = \sum_{k \in A_i} 2p_i +
q_i$.  Note that $\gamma_i \ge 2\E[|A_i \cap \Delta|]$
for one-sided routing and 
$\gamma_i \ge 2\E[|A_i \cap \Delta|] + \E[|A_i \cap \beta|]$ 
for two-sided routing.  
Observe also that
$\sum_{i=0}^{\infty} \gamma_i$ is at most
$2\ell$ for one-sided routing and at most $2\ell + \ell^2$ for two-sided
routing.

Consider some $S=S^t$.  
Let $A$ be the event $\left[\lndrop < \ln a\right]$.
If $|S| \ge a$,
then by Lemma~\ref{lemma-log-drop} we have
\begin{equation}
\label{eq-log-drop-revisited}
\E \left[ \lndrop : S^t, A \right]
\le \constdrop + \frac{\ln \E\left[1+Z : S^t\right]}{\Pr[A: S^t]},
\end{equation}
where
$Z = 2|\Delta \cap S'|$ with one-sided routing and
$Z = 2|\Delta \cap S'| + |\beta \cap S'|$ with two-sided routing,
with $S' = [\min(S) + \ceil{\aS} -1 , \max(S)-1]$ in each case, as in
Lemma~\ref{lemma-log-drop}.

As we observed earlier, our choice of $a$ and
Lemma~\ref{lemma-aggregate-max-drop} imply
$\Pr[\lndrop \ge \ln a : S^t] \le \ln^{-3} n$, so 
$\Pr[A: S^t] = 1-\Pr[\lndrop \ge \ln a: S^t] 
\ge 1 - \ln^{-3} n \ge \frac{1}{2}$ for sufficiently
large $n$.
So we can replace (\ref{eq-log-drop-revisited}) with
\begin{equation}
\label{eq-log-drop-revisited-simple}
\E \left[ \lndrop : S^t, A \right]
\le \constdrop + 2 \ln \E\left[1+Z : S^t\right],
\end{equation}

Let us now obtain a bound on $\ln \E[1+Z]$ in terms of $|S|$ and
the $p_i$ and $q_i$.
For one-sided routing, we use the fact that $|S| > 1$ implies
$S=\{1\ldots|S|\}$.  For two-sided routing, we use monotonicity of the
$p_i$ and $q_i$ to replace $S$ with $\{1\ldots|S|\}$;
in particular, to replace a sum of $2p_i+q_i$ over a subrange of $S$
with a sum over subrange of $\{1\ldots|S|\}$ that is at least as
large. 
In either case, we get that
\begin{equation}
\label{eq-bound-one-plus-z}
\ln \E[1+Z] \le \ln\left(1 + \sum_{i = \ceil{\aS}-1}^{|S|-1} 2p_i + q_i\right),
\end{equation}
and thus
$\E \left[ \lndrop : S^t, A \right]$ is bounded by
\begin{equation}
\label{eq-lower-bound-mu}
\mu_{\ln |S|} = 
\constdrop + 
2\ln\left(1 + \sum_{i = \ceil{\aS}-1}^{|S|-1} 2p_i + q_i\right),
\end{equation}
provided $|S| \ge a$.
For $|S| < a$, set $\mu_{\ln|S|} = \ln a$.

\newcommand{\gzs}{\gamma_{z'} + \gamma_{z'+1} + \gamma_{z'+2}}
\begin{sloppypar}
Let us now compute $m_z$, as defined in (\ref{eq-mean-lower-bound-m}).
For $z < \ln a$, $m_z = \ln a$.
For larger $z$, observe that
$m_z = \sup \left\{ m_{\ln |S|} : e^z \le |S| < a e^z \right\}$.
Now if $e^z \le |S| < a e^z$, then the bounds on the sum in
(\ref{eq-lower-bound-mu}) both lie between $\ceil{a^{-1} e^z}-1$ and
$a e^z -1$, so that
\begin{eqnarray*}
\label{eq-lower-bound-m}
m_z &\le& 
\constdrop +
2\ln\left(1 + \sum_{i = \ceil{a^{-1}e^z}-1}^{\floor{ae^z-1}} 2p_i + q_i\right)
\\
&\le&
\constdrop +
2\ln(1 + \gzs),
\end{eqnarray*}
where $z' = \floor{z/\ln a} - 1$.
\end{sloppypar}

Finally, compute
\begin{eqnarray*}
T(\ln n)  &=&
\int_{0}^{\ln n} \frac{1}{m_z} dz \\
&\ge&
\int_{\ln a}^{\ln n} \frac{1}{\constdrop+2\ln(1+\gzs)} dz \\
&\ge&
  \sum_{i = 0}^{\floor{\ln n / \ln a} - 1}
    \frac{\ln a}{\constdrop+2\ln(1+\gamma_i + \gamma_{i+1} + \gamma_{i+2})}.
\end{eqnarray*}

To get a lower bound on the sum, 
note that
\[\sum_{i = 0}^{\floor{\ln n / \ln a} - 1}
  (\gamma_i + \gamma_{i+1} + \gamma_{i+2})
 \le 3 \sum_{i=0}^{\floor{\ln n / \ln a} + 1} \gamma_i
 \le 3 \sum_{i=0}^{\infty} \gamma_i,
 \]
which is at most $L = 6\ell$ for one-sided routing and at most
$L = 6\ell+3\ell^2$ for two-sided routing.
In either case, because $\frac{1}{c+2\ln(1+x)}$ is convex and decreasing,
we have
\begin{eqnarray}
T(\ln n) &\ge&
  \sum_{i = 0}^{\floor{\ln n / \ln a} - 1}
    \frac{\ln a}{\constdrop + 2\ln(1+\gamma_i + \gamma_{i+1} + \gamma_{i+2})}
\nonumber\\
&\ge&
\sum_{i = 0}^{\floor{\ln n / \ln a} - 1}
    \frac{\ln a}{\constdrop+2\ln\left(1 + \frac{L}{\floor{\ln n / \ln a}}\right)}
\nonumber\\
&=&
\frac{ \ln a
    \floor{\ln n / \ln a}
}{
    \constdrop +
    2\ln\left(1 + \frac{L}{\floor{\ln n / \ln a}}\right)}.
    \label{eq-mean-lower-bound-ugly-T}
\end{eqnarray}

We will now rewrite our bound on $T(\ln n)$ in a more convenient asymptotic
form.  We will ignore the $1$ and concentrate on the large fraction.
Recall that $a = 3 \ell \ln^3 n$,
so $\ln a = \Theta(\ln \ell + \ln \ln n)$.
Unless $\ell$ is polynomial in $n$, we have $\ln n / \ln a =
\omega(1)$ and the numerator simplifies to $\Theta(\ln n)$.

Now let us look at the denominator.
Consider first the term $\constdrop$.
We can rewrite this term as $-\ln(1-a^{-1})$; since $a^{-1}$ goes to
zero as $\ell$ and $n$ grow we have 
$-\ln(1-a^{-1}) = \Theta(a^{-1}) = \Theta(\ell^{-1} \ln^{-3} n)$.
It is unlikely that this term will contribute much.

\begin{sloppypar}
Turning to the second term, let us use the fact that 
$\ln(1+x) \le x$ for $x \ge 0$.
Thus
\begin{eqnarray*}
2\ln\left(1+\frac{L}{\floor{\ln n / \ln a}}\right)
&\le& 2\,\frac{L}{\floor{\ln n/\ln a}}\\
&=& O\left(\frac{L(\log l + \log \log n)}{\log n}\right),
\end{eqnarray*}
and the bound in (\ref{eq-mean-lower-bound-ugly-T}) simplifies to
$\Omega\left(\log^2 n / \left( L (\log \ell + \log \log n)\right)\right)$.
We can further assume that $\ell = O(\log^2 n)$, since otherwise the
bound degenerates to $\Omega(1)$, and
rewrite it simply as $\Omega\left(\log^2 n / \left(L \log \log n\right)\right).$
\end{sloppypar}

For large $L$, the approximation
$\ln(1+x) \le 1+\ln x$ for $x \ge 0.59$ is more useful.
In this case (\ref{eq-mean-lower-bound-ugly-T}) simplifies to 
$T(\ln n) = \Omega(\ln n / \ln \ell)$, which has a natural
interpretation in terms of the tree of successor nodes of some single
starting node and gives essentially the same bound as 
Theorem~\ref{theorem-tree-lower-bound}. 

We are not quite done with Theorem~\ref{theorem-mean-lower-bound} yet,
as we still need to plug our $T$ and $\epsilon$ into
(\ref{eq-mean-lower-bound}) to get a lower bound on $\E[\tau]$.
But here we can simply observe that 
$\epsilon T = O(1/\log n)$, so the denominator in
(\ref{eq-mean-lower-bound}) goes rapidly to $1$.
Our stated bounds are thus finally obtained by substituting $O(\ell)$
or $O(\ell^2)$ for $L$.
\end{proof}

\subsubsection{Possible strengthening of the lower bound}

Examining the proof of Theorem~\ref{theorem-lower-bound}, 
both the $\ell^2$ that appears in the bound
(\ref{eq-lower-bound-two-sided}) for two-sided
routing and the extra conditions imposed on the $\Delta$ distribution
arise only as artifacts of our need to project each range $S$ onto
$\{1\ldots|S|\}$ and thus reduce the problem to tracking a single
parameter.  We believe that a more sophisticated argument that does
not collapse ranges together would show a stronger result:
\begin{conjecture}
Let $G$, $\Delta$, and $\ell$
be as in Theorem~\ref{theorem-lower-bound}.
Consider a greedy routing trajectory starting at a point chosen
uniformly from $1 \ldots n$ and ending at $0$.

Then the expected time to reach $0$ is
\begin{displaymath}
\Omega\left(
      \frac{\log^2 n}{\ell \log \log n}
\right),
\end{displaymath}
with either one-sided or two-sided routing, and no constraints on the
$\Delta$ distribution.
\end{conjecture}

We also believe that the bound continues to hold in higher dimensions
than $1$.  Unfortunately, the fact that we can embed the line in, say,
a two-dimensional grid is not enough to justify this belief;
divergence to one side or the other of the line may change the
distribution of boundaries between segments and break the proof of
Theorem~\ref{theorem-lower-bound}.
\subsection{Upper Bounds}
\label{sec:UPPERBNDS}

In this section, we present upper bounds on the 
delivery time of messages in a simple metric 
space: a one-dimensional real line. To simplify
theoretical analysis, the system 
is set up as follows.
\begin{itemize}
  \item Nodes are embedded at grid points on the real
        line.
  \item Each node $u$ is connected to its nearest 
        neighbor on either side and to one or more
        long-distance neighbors.
  \item The long-distance neighbors are chosen as
        per the inverse power-law distribution with
        exponent $1$, i.e.,
        each long-distance neighbor $v$ is chosen 
        with probability inversely proportional to 
        the distance between $u$ and $v$. Formally,
        Pr[$v$ is the $i$th neighbor of $u$] = 
        $(\frac{1}{d(u,v)})/(\sum_{v'\neq u}\frac{1}{d(u,v')})$,
        where $d(u,v)$ is the distance between nodes
        $u$ and $v$ in the metric space.
  \item Routing is done greedily by forwarding the 
        message to the neighbor closest to the target 
        node.
\end{itemize}

We analyze the performance for the cases of a single 
long-distance link and of multiple ones, both in a failure-free network
and in a network with link and node failures. Note that when
we say {\em node}, we actually refer to a vertex in the 
virtual overlay network and not a {\em physical} node as 
in the earlier sections.

\subsubsection{Single Long-Distance Link}
\label{sec:INVERSE}

We first analyze the delivery time in an idealized model with no 
failures and with one long-distance link per node.
Kleinberg \cite{KL99} proved that with $n^d$ nodes embedded at grid 
points in a $d$-dimensional grid, with each node $u$ connected to its 
immediate neighbors and one long-distance neighbor $v$ chosen with 
probability proportional to $1/d(u, v)^d$, any message can be 
delivered in time polynomial in $\log n$ using greedy routing. 
While this result can be directly applied to our model with $d=1$
and $l=1$ to give a $O(\log^2 n)$ delivery time, we get a much simpler 
proof by use of Lemma~\ref{lemma-probabilistic-recurrence-ub}. 
We include the proof below for completeness.

\begin{theorem}
\label{thm:UPPER-SINGLE}
Let each node be connected to its immediate neighbors (at distance 1)
and $1$ long-distance neighbor chosen with probability inversely 
proportional to its distance from the node. Then the expected delivery 
time with $n$ nodes in the network is $T(n)=O(H_n^2)$.
\end{theorem}

\begin{proof}
Let $\mu_k$ be the expected number of nodes crossed when the message is 
at a node that is at a distance $k$ from the destination. 
Clearly, $\mu_k$ is non-decreasing.

\begin{figure}[htb]
\centerline{\epsfig{figure=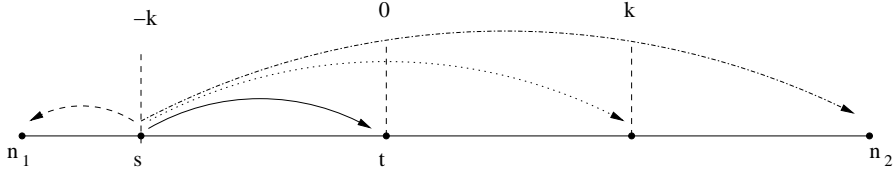, height=75pt}}
\caption{All the possible distances that can be
covered from source node $s$.}
\end{figure}

\noindent
Let
$$\mu_k = \frac{\sum_{i=1}^k \frac{1}{i} \cdot i}{S}
        + \frac{\sum_{i=1}^{k-1} \frac{1}{2k-i} \cdot i}{S}
        + \frac{\sum_{i=1}^{n_1-k} \frac{1}{i} \cdot 1}{S}
        + \frac{\sum_{i=2k}^{n_2+k} \frac{1}{i} \cdot 1}{S},$$
where
$$
S = \sum_{i=1}^{n_1-k} \frac{1}{i} + \sum_{i=1}^{n_2+k} \frac{1}{i}\\
  = H_{n_1-k}+H_{n_2+k} < 2H_n.
$$
Then
$$
\mu_k > \frac{1}{S} [ k + 0 + H_{n_1-k} + H_{n_2+k} - H_{2k}]\\
      > \frac{k}{S} > \frac{k}{2H_n}.\\
$$
Clearly, $\mu_k$ is non-decreasing, and thus
using Lemma~\ref{lemma-probabilistic-recurrence-ub}, we get
$$T(n) \leq \sum_{k=1}^n \frac{1}{\mu_k} 
= \sum_{k=1}^n \frac{2H_n}{k}= O(H_n^2).$$
Thus with this distribution, the delivery time is 
logarithmic in the number of nodes. 
\end{proof}

\subsubsection{Multiple Long-Distance Links}
\label{sec:MULT-LINKS}

The next interesting question is whether we can improve the $O(\log^2 n)$
delivery time by using multiple links instead of a single one. In 
addition to improvement in performance, multiple links  also give the
benefit of robustness in the face of failures. We first look at 
improvement in performance by using multiple links in the system
and then go onto analysis of failures in Section~\ref{sec:LINK-FAIL}.
Suppose that there are $\ell$ links from each node. 
We consider different strategies for generating links and routing
depending on number of links $\ell$ in two ranges: $\ell \in [1,\lg n]$ 
and $\ell \in (\lg n, n^c]$. 

In \cite{KL01}, Kleinberg uses a group structure to get a delivery time
of $O(\log n)$ for the case of a polylogarithmic number of links.
However, he uses a more complicated algorithm for routing while we 
obtain the same bound (for the case of a line) using only greedy routing. 

\begin{figure}[h]
\begin{center}
\input{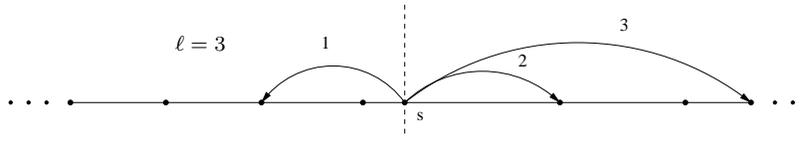}
\caption{Multiple long-distance links for each node.}
\end{center}
\end{figure}

\paragraph{Upper Bound}
Let us first consider a randomized strategy for link distribution 
when $\ell \in [1, \lg n]$.

\begin{theorem}
\label{thm:UPPER-RANDOMIZED-MULTIPLE}
Let each node be connected to its immediate neighbors (at distance 1)
and $\ell$ long-distance neighbors chosen independently with replacement
with probability proportional to their distances from the node.
Let $\ell \in [1, \lg n]$.  Then the expected delivery time 
$T(n)=O(\log^2 n/\ell)$.
\end{theorem}

\begin{proof}
The basic idea for this proof comes from Kleinberg's model~\cite{KL99}.
Kleinberg considers a two-dimensional grid with nodes at every grid point.
The delivery of the message is divided into phases. A message is said to 
be in phase $j$ if the distance from the current node to the destination 
node is between $2^j$ and $2^{j+1}$. There are at most ($\lg n+1$) such 
phases. He proves that the expected time spent in each phase is at most
$O(\log n)$, thus giving a total upper bound of $O(\log^2 n)$ on the delivery 
time. We use the same phase structure in our model, and this proof 
is along similar lines.

In our multiple-link model, each node has $\ell$ long-distance neighbors
chosen with replacement.  The probability that $u$ chooses a node $v$ as its 
long-distance neighbor is
$1-(1-q)^\ell$, where $q=\frac{d(u, v)^{-1}}{\sum_{u\ne v}d(u, v)^{-1}}$.
We can get a lower bound on this probability as follows:
\begin{eqnarray*}
1-(1-q)^\ell &>& 1 - (1 - q\ell + \frac{\ell(\ell-1)}{2}q^2)\\
&=&q\ell - \frac{\ell(\ell-1)}{2}q^2 = q\ell\left[1-\frac{(\ell-1)q}{2}\right]\\
&=&q\ell\left[1 -\frac{\ell q}{2} +\frac{q}{2}\right]\\
&\geq&q\ell\left[1-\frac{\ell q}{2}\right].\\
\end{eqnarray*}

Notice that $\ell q < 1$, because $q < \frac{1}{\lg n}$ and $\ell \leq \lg n$.
So, the probability that $u$ chooses $v$ as its long-distance 
neighbor is at least 

\begin{eqnarray*}
q\ell\left[1-\frac{\ell q}{2}\right]
&\geq&q\ell\left[1-\frac{1}{2}\right]=\frac{q\ell}{2}
=\ell [2d(u,v)H_n]^{-1}.
\end{eqnarray*}

Now suppose that the message is currently in phase $j$. 
To end phase $j$ at this step, the message should enter a set of nodes $B_j$
at a distance $\leq 2^j$ of the destination node $t$. There are at least $2^j$
nodes in $B_j$, each within distance $2^{j+1} + 2^j < 2^{j+2}$ of $u$. So the
message enters $B_j$ with probability 
$\geq 2^j\ell\frac{1}{2H_n2^{j+2}} = \frac{\ell}{8H_n}$

Let $X_j$ be the total number of steps spent in phase $j$. Then
$$
E[X_j] = \sum_{i=1}^\infty Pr[X_j \geq i]
\leq \sum_{i=1}^\infty\left( 1 - \frac{\ell}{8H_n} \right)^{i-1}
= \frac{8H_n}{\ell}.
$$

Now if $X$ denotes the total number of steps, then
$X=\sum_{j=0}^{\lg n}X_j$, and by linearity of expectation, we get 
$EX\leq(1+\lg n)(8H_n/\ell)=O(\log^2n/\ell)$.
\end{proof}

For $\ell \in (\lg n, n^c]$, we use a deterministic strategy. We represent 
the location of each node as a number in a base $b\geq 2$, and 
generate links to nodes at distances $1x, 2x, 3x, \ldots, (b-1)x$, for each 
$x \in \{b^0, b^1, \ldots, b^{\lceil\log_b n\rceil -1} \}$.
Routing is 
done by eliminating the most significant digit of the distance at each step. 
As this distance can be at most $b^{\lceil\log_b n\rceil}$, we get 
$T(n)=O(\log_b n)$. This strategy is similar in spirit to Plaxton's
algorithm \cite{PL97}.

Some special cases are instructive.
Let $\ell=O(\log n)$ and let each node link to nodes in both directions at 
distances $2^i, 1 \leq i \leq 2^{\log n-1}$, provided nodes are present at
those distances. This gives $T(n)=O(\log n)$. Similarly let $\ell=O(\sqrt{n})$. 
Links are established in both directions to existing nodes at distances $1, 2, 
3, \ldots, \sqrt{n}, 2\sqrt{n}, 3\sqrt{n}, \ldots, \sqrt{n}(\sqrt{n}-1)$, 
giving $T(n)=O(1)$. In fact, $T(n)=O(1)$ when $b={n^c}$, for any fixed $c$.

\begin{theorem}
\label{thm:UPPER-BOUND-DETERMINISTIC-MULTIPLE}
Choose an integer $b>1$. With $\ell=(b-1)\lceil\log_b n\rceil$, let 
each node link 
to nodes at distances $1x, 2x, 3x, \ldots, (b-1)x$, for each $x \in \{b^0, b^1, 
\ldots, b^{\lceil \log_b n\rceil -1} \}$. Then the delivery time $T(n) 
 = O(\log_b n)$.
\end{theorem}

\begin{proof}
Let $d_1, d_2, \ldots d_t$ be the distances of the successive nodes in the 
delivery path from the target $t$, where  $d_1$ is the distance of the source node 
and $d_t=0$.  For each $d_i, \exists k_i \in \{0, 1, \ldots, 
\lfloor \log_b n\rfloor\}$ such that 
$$b^{k_i} \leq d_i < b^{k_i+1}.$$
Hence
$$1 \leq \lfloor \frac{d_i}{b^{k_i}} \rfloor < b.$$
Now each node is connected to the node at distance $b^{k_i} \lfloor 
\frac{d_i}{b^{k_i}} \rfloor$. We get
$$
d_{i+1} = d_i - b^{k_i} \lfloor \frac{d_i}{b^{k_i}} \rfloor
        = d_i\mod b^{k_i}
        < b^{k_i}.
$$
Thus $k_i$ drops by at least 1 at every step. As $k_1 \leq \lceil
\log_b n\rceil$, we get 
$T(n)=O(\log_b n)$.
\end{proof}

\subsubsection{Failure of Links}
\label{sec:LINK-FAIL}

It appears that our
linking strategies may fail to give the same delivery time
in case the links fail. However, we show that we get reasonable
performance even with link failures.  In our model, we assume
that each link  is present independently with probability $p$.
Let us first look at
the randomized strategy for number of links $\ell \in [1, \lg n]$.\\

\begin{figure}[htb]
\centerline{\epsfig{figure=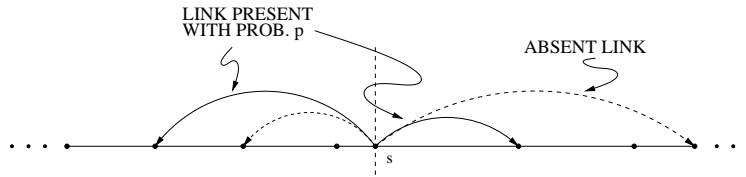, height=75pt}}
\caption{Each long-distance link is present with probability $p$.}
\end{figure}

Our proof is along similar
lines as our proof for the case of no failures. 
Intuitively, since some of the links fail, we
expect to spend more time in each phase and this time 
should be inversely proportional to the probability with
which the links are present. We prove that the expected time
spent in one phase is $O(\log n/p\ell)$, which gives a total
delivery time of $O(\log^2 n/p \ell)$. We assume that the links
to the immediate neighbors are always present so that a message
is always delivered even if it takes very long.

\begin{theorem}
Let the model be as in Theorem~\ref{thm:UPPER-RANDOMIZED-MULTIPLE}.
Assume that the links to the immediate neighbors are always present.
If the probability of a long-distance link being present is $p$,
then the expected delivery time is $O(\log^2 n/p\ell)$.
\end{theorem}

\begin{proof}
Recall that in case of no link failures, the probability that $u$ 
chooses a node $v$ as its long-distance neighbor is at 
least $q\ell/2$ 
where $q=\frac{d(u, v)^{-1}}{\sum_{u\ne v}d(u, v)^{-1}}$.

Now when we consider link failures, given that $u$ chose
$v$ as its long-distance neighbor, the probability that
there is a link present between $u$ and $v$ is $p$.
So, the probability that $u$ chooses a node $v$ as its long-distance neighbor 
is at least $pq\ell/2 = p\ell[2d(u,v)H_n]^{-1}$.

The rest of the proof is the same as the proof for 
theorem~\ref{thm:UPPER-RANDOMIZED-MULTIPLE}. Let $X_j$ be the
number of steps spent in phase $j$. Then
$$E[X_j]=\sum_{i=1}^\infty Pr[X_j \geq i]
= \frac{8H_n}{p\ell}.$$

If $X$ denotes the total number of steps, then by linearity of
expectation, we get 
$EX\leq(1+\lg n)(8H_n/p\ell)=O(\log^2n/p\ell)$.
\end{proof}

We turn to the deterministic strategy with $\ell \in (\lg n, n^c]$
links. A similar intuition works for $\ell \in (\lg n, n^c]$. If a 
link fails, then the node has to take a shorter long-distance link,
which will not take the message as close to the target as the initial 
failed link. Clearly as $p$ decreases, the message has to take
shorter and shorter links which increases the delivery time.

To make the analysis simpler, we
change the link model a bit and let each node be
connected to other nodes at distances $b^0, b^1, b^2, \ldots, 
b^{\lfloor \log_b n \rfloor}$.
Once again, we compute the expected distance covered from the
current node and use Lemma~\ref{lemma-probabilistic-recurrence-ub}
to get a delivery time of $O(b \log n/p)$.  As $p$ decreases,
the delivery time increases; whereas as $b$ decreases,
the delivery time decreases, but the
information stored at each node increases.

\begin{theorem}
Let the number of links be $O(\log_b n)$, and let each node have a link
to distances $b^0, b^1, b^2, \ldots, b^{\lfloor \log_b n \rfloor}$.
Assume that the links to
the nearest neighbors are always present. If the probability of 
a link being present is $p$, then the delivery time 
$T(n)= O(bH_n/p)$.
\end{theorem}

\begin{proof}
Let the distance of the current node
from the destination be $k$. Let $\mu_k$ represent the distance covered 
starting from this node. Then with probability $p$, there will be a 
link covering distance $\flrbk{}$. If this link is absent with 
probability $q=1-p$, then we can cover a distance $\flrbk{-1}$ 
with a single link with probability $pq$ and so on. In general,
the average distance $\mu_k$ covered when the message is at distance $k$ 
from the destination is
\begin{eqnarray*}
\mu_k&=&p\flrbk{} + pq\flrbk{-1} + \ldots 
           + pq^{\lfloor \log_b k \rfloor-1}b^1
           + q^{\lfloor \log_b k \rfloor}b^0 \\
&\geq& \sum_{i=0}^{\lfloor \log_b k \rfloor}
       p\flrbk{-i}q^i\\
&=&p\flrbk{} \sum_{i=0}^{\lfloor \log_b k \rfloor} \left(\frac{q}{b}\right)^i\\
&=&p\flrbk{} \frac{1-\left(q/b\right)^{\lfloor \log_b k \rfloor+1}}{1-(q/b)}\\
&=&\frac{p(\flrbk{+1}-q^{\lfloor \log_b k \rfloor+1})}{b-q}\\
&\geq&\frac{p(bk/b-1)}{b-q}\\
&\geq&\frac{p(k-1)}{2(b-q)}.\\
\end{eqnarray*}
Using Lemma~\ref{lemma-probabilistic-recurrence-ub}, we get
$$
T(n) \leq \sum_{k=1}^n\frac{1}{\mu_k}
=1+\sum_{k=2}^n\frac{2(b-q)}{p(k-1)}
=1+\frac{2(b-q)}{p}\left[\sum_{k=2}^n\frac{1}{(k-1)}\right]
=O(bH_n/p).
$$
\end{proof}

\subsubsection{Failure of Nodes}
\label{sec:NODE-FAIL}

We consider two different cases of node failures when  we study 
their effect on system performance. In the first case, as described in 
Section~\ref{sec:BIN-NODE-FAIL}, some of the nodes may fail 
and then the remaining nodes will link to each other as 
per the link distribution. In the second case, as explained
in Section~\ref{sec:GEN-NODE-FAIL}, the nodes first link to 
their neighbors and then some of the nodes may fail. 

\paragraph{Binomially Distributed Nodes}
\label{sec:BIN-NODE-FAIL}

Let $p$ be the 
probability that a node is present at any point. Here also, each node is
connected to its nearest neighbors and one long-distance neighbor. In
addition, the probability of choosing a particular node as a long-distance 
neighbor is conditioned on the existence of that node. 

\begin{theorem}
\label{thm:UPPER-BINOMIAL}
Let the model be as in Theorem~\ref{thm:UPPER-SINGLE}.
Let each node be present with probability $p$ and all nodes
link only to existing nodes. Then the worst-case expected delivery time 
is $O(\log^2 n)$.
\end{theorem}

\begin{proof}
We bound the expected drop $\mu_k$ as follows:

\begin{eqnarray*}
\mu_k &=& \frac{\sum_{i=1}^k \frac{1}{i} \cdot i \cdot p}{p \cdot S}
       +  \frac{\sum_{i=1}^{k-1} \frac{1}{2k-i} \cdot i \cdot p}{p \cdot S}
       +  \frac{\sum_{i=1}^{n_1-k} \frac{1}{i} \cdot 1 \cdot p}{p \cdot S}
       +  \frac{\sum_{i=2k}^{n_2+k} \frac{1}{i} \cdot 1 \cdot p}{p \cdot S}\\
      &>& \frac{1}{S} [ k + 0 + H_{n_1-k} + H_{n_2+k} - H_{2k}]\\
      &>& \frac{k}{S} > \frac{k}{2H_n}.\\
\end{eqnarray*}

Using Lemma~\ref{lemma-probabilistic-recurrence-ub}, 
we get $T(n)\leq \sum_{k=1}^n 1/\mu_k
=O(H_n^2)$. This is exactly the same result that we get in
Section~\ref{sec:INVERSE} where all the nodes are present. 
\end{proof}

This result is not
surprising because if nodes link only to other existing nodes, the
only difference is that we get a smaller random graph. This does
not affect the routing algorithm or the delivery time.

\paragraph{General Failures}
\label{sec:GEN-NODE-FAIL}

We observe that the analysis for node failures is not as simple as that
for link failures because we no longer
have the important property of independence that we have 
in the latter case. In the case of link failures, 
the nodes first choose their neighbors and then it is possible that
some of these links fail; thus, the event that a node is connected
to another node is completely independent of the event that, say, its
neighbor is connected to the same node. Each link fails independently, and
so the accessibility of a target node from any other node depends only
on the presence of the link between the two nodes in question.

In case of node failures, this important independence
property is no longer true. Suppose that a
node $u$ cannot communicate with some other node $v$ (because $v$
failed), even though there may be a functional link between $u$
and $v$. Now the probability of some other node $w$ being able
to communicate with $v$ is not independent of the probability 
that $u$ can communicate with $v$ because the probability of 
$v$ being absent is common for both the cases. This complicates 
the analysis of the performance because it is no longer the case 
that if one node cannot communicate with some other node, it has a 
good chance of doing so by passing the message to its neighbor. 

In order to analyze this situation, we consider jumps only to one 
phase lower rather than jumping over several phases.  The idea is 
that the jumps between phases are independent, so once we move from 
phase $j$ to phase $j-1$, further routing no longer depends on
any nodes in phase $j$. We can condition on the number of nodes being 
alive in the lower phase and estimate the time spent in each phase. 
Intuitively, if a node is present with probability $p$, we would expect 
to wait for a time inversely proportional to $p$ in anticipation of 
finding a node in the lower phase to jump to.

\begin{theorem}
Let the model be as in Theorem~\ref{thm:UPPER-RANDOMIZED-MULTIPLE}
and let each node fail with probability $p$.
Then the expected delivery time is $O(\log^2n/(1-p))$.
\end{theorem}

\begin{proof}
Let $T$ be the time taken to drop down from layer
$j$ to layer $j-1$. Let $l$ out of $N$ nodes be alive 
in layer $j-1$ and let $q$ be the probability that
a node in layer $j$ is connected to some node in 
layer $j-1$. Then the expected time to drop to
layer $j-1$, given that there are $l$ live nodes
in it, is given by
\begin{eqnarray*}
E[T|l] &=& 1 + \left[ (1-q) + \frac{q(N-l)}{N} \right] E[T|l]\\
&=& \frac{N}{ql}.
\end{eqnarray*}

Now $l$ can vary between $1$ and $N$. (Note that $l$
cannot be $0$ because if there are no live nodes in the
lower layer, the routing fails at this point.) 
We get
\begin{eqnarray*}
E[T] &=& 
\sum_{l=1}^{N}\frac{N}{ql}\left[ p^{N-l}(1-p)^l{{N}\choose{l}} \right]\\
&=&\frac{N}{q}\sum_{l=1}^{N}\frac{1}{l}p^{N-l}(1-p)^{l}{N\choose l}\\
&\leq&\frac{N}{q}\sum_{l=1}^{N}\frac{2}{l+1}p^{N-l}(1-p)^{l}{N\choose l}\\
&=&\frac{2N}{q(N+1)(1-p)}\sum_{l=1}^{N}p^{N-l}(1-p)^{l+1}{N+1\choose l+1}\\
&\leq&\frac{2N}{q(N+1)(1-p)}\left[p+(1-p)\right]^{N+1}\\
&=&\frac{2N}{q(N+1)(1-p)}.
\end{eqnarray*}

Not surprisingly, the expected waiting time in a layer
is inversely proportional to the probability of being
connected to a node in the lower layer and to the probability
of such a node being alive.

For our randomized routing strategy with $[1, \lg n]$ links,
$q\approx 1/(H_n\ell)$. Since there are at most $(\lg n +1)$ layers, 
we get an expected delivery time of $O(\log^2 n/(1-p)\ell)$.
\end{proof}

In contrast, for our deterministic routing strategy, certain 
carefully chosen node failures can lead to dismal situations where a
message can get stuck in a local neighborhood with no hope of getting 
out of it or eventually reaching the destination node. We conjecture 
that this should be a very low probability event, so its occurrence 
will not affect the delivery time considerably. We have not yet analyzed
this situation formally.

\section{Construction of Graphs}
\label{sec:RANDOMGRAPHS}

As the group of nodes present in the network changes, so does the
graph of the virtual overlay network. In order for our routing  
techniques to be effective, the graph must always exhibit the 
property that the likelihood of any two vertices $v,u$ being connected
is $\Omega(1/d(v,u))$. We describe a heuristic approach to 
construct and maintain a random graph with such an invariant.

Since the choice of links leaving each vertex is independent of 
the choices of other vertices, we can assume that points
in the metric space are added one at a time. Let $v$ be the $k$-th
point to be added. Point $v$ chooses the sinks of its outgoing links
according to the inverse power law distribution with exponent $1$ 
and connects to them by
running the search algorithm. If a desired sink $u$ is not present, $v$
connects to $u$'s closest live neighbor. In effect, each of the 
$k-1$ points already present before $v$ is surrounded by a basin of 
attraction, collecting probability mass in proportion to its length. 
Since we assume the hash function populates the metric space evenly, 
and because of absolute symmetry, the basin length $L$ has the same 
distribution for all points. It is easy to see that with high probability, 
$L$ will not be much smaller than its expectation: $\prob{L \leq c 
\cdot k^{-1}}=1-(1-c\cdot k^{-1})^{k-1}$. A lower bound on the 
probability that the link $(v,u)$ is present is $c' \cdot k^{-1} 
\cdot d(v,f)^{-1}$, where $f$ is the point in $u$'s basin that is the 
farthest from $v$.\footnote{The constant $c'$ has absorbed $c$ and the 
normalizing constant for the distribution.} However, the bound holds 
only if $u$ is among the $k-1$ points added before $v$. Otherwise, 
the aforementioned probability is $0$, which means that we need to amend 
our linking strategy to transfer probability mass from the case of 
$u$ having arrived before $v$ to the case of $u$ having arrived after $v$.
We describe next how to accomplish this task.

Let $v$ be a new point.  We give earlier points the opportunity
to obtain outgoing links to $v$ by having $v$ (1) calculate the
number of incoming links it ``should'' have from points added before it 
arrived, and (2) choose such points according to the inverse 
power-law distribution with exponent 1.\footnote{All this can be easily 
calculated by $v$ since the link probabilities are symmetric.} If $\ell$ 
is the number of outgoing links for each point, then $\ell$ will also be 
the expected number of incoming links that $v$ has to estimate in step 
(1).  
We approximate the number of links 
ending at $v$ by using a Poisson distribution with 
rate $\ell$, that is, the probability that $v$ has $k$ incoming links is 
$\frac{e^{-l}l^k}{k!}$, and the expectation of the distribution is $\ell$. 

After step (2) is completed by $v$, each chosen point $u$ responds to 
$v$'s request by choosing one of its existing links to be replaced by 
a link to $v$. The choice of the link to replace can vary. We use a 
strategy that 
builds on the work of Sarshar~\etal~\cite{SarsharR02}. In that work, the 
authors use ideas of Zhang~\etal~\cite{ZhangGG02} to build a graph where each 
node has a single long-distance link to a node at distance $d$ with probability 
$1/d$. When a node with a long-distance link at distance $d_1$ encounters a 
new node at distance $d_2$, either due to its arrival or due to a data request, 
it replaces its existing link with probability $p_2/(p_1+p_2)$, where
$p_i=1/d_i$, and links to the new node. We extend this idea to our case of 
multiple long-distance links. Consider a node $u$ with $k$ neighbors at distances 
$d_1, d_2, \ldots, d_k$. When a new node $v$ at distance $d_{k+1}$ 
requests an incoming link from $u$, $u$ replaces one of its existing links
with a link to $v$ with probability $p_{k+1}/\sum_{j=1}^{k+1}p_j$. This is 
a trivial extension of the formula $p_2/(p_1+p_2)$ of \cite{SarsharR02}.
However, this probability must now be distributed among $u$'s $k$ existing long-distance
links since $u$ needs to choose one of them to redirect to $v$. We choose to
do that according to the inverse power-law distribution with exponent 1, that is,
$u$ chooses to replace its link to the node at distance $d_i$, $1\leq i \leq k$, 
with probability $(p_i/\sum_{j=1}^{k} p_j)$. Hence, the probability that $u$ 
decides to link to $v$ and decides to replace its existing link to the node at 
distance $d_i$ with a link to $v$ is equal to $(p_i/\sum_{j=1}^{k} p_j) \cdot
(p_{k+1}/\sum_{j=1}^{k+1}p_j)$. Notice that $u$ may decide not to redirect 
any of its existing links to $v$ with probability $1-p_{k+1}/\sum_{j=1}^{k+1}p_j$.
The intuition for using such replacement strategy comes from the invariant that we
want to maintain dynamically as new nodes arrive: $u$ has a link to a node 
$i$ at distance $d_i$ with probability inversely proportional to $d_i$; hence, 
conditioning on $u$ having $k$ long-distance links, the following equation must hold.
\begin{eqnarray*}
\prob{\mbox{$u$ replaces link to $i$ with link to $v$}} & = &
\prob{\mbox{$u$ has a link to $i$ before $v$ arrives}} \\
& - & \prob{\mbox{$u$ has a link to $i$ after $v$ arrives}} \\
& = & \frac{p_i}{\sum_{j=1}^k p_j} - \frac{p_{i}}{\sum_{j=1}^{k+1} p_j} \\
& = & \frac{p_i}{\sum_{j=1}^k p_j} \cdot \frac{p_{k+1}}{\sum_{j=1}^{k+1}p_j}. \\
\end{eqnarray*}
The same heuristic can be used for regeneration of links when a node crashes.

\begin{figure}
\centering
  \mbox{\subfigure[The derived distribution.
    \label{fig:DISTRIBUTION}]
       {\epsfig{figure=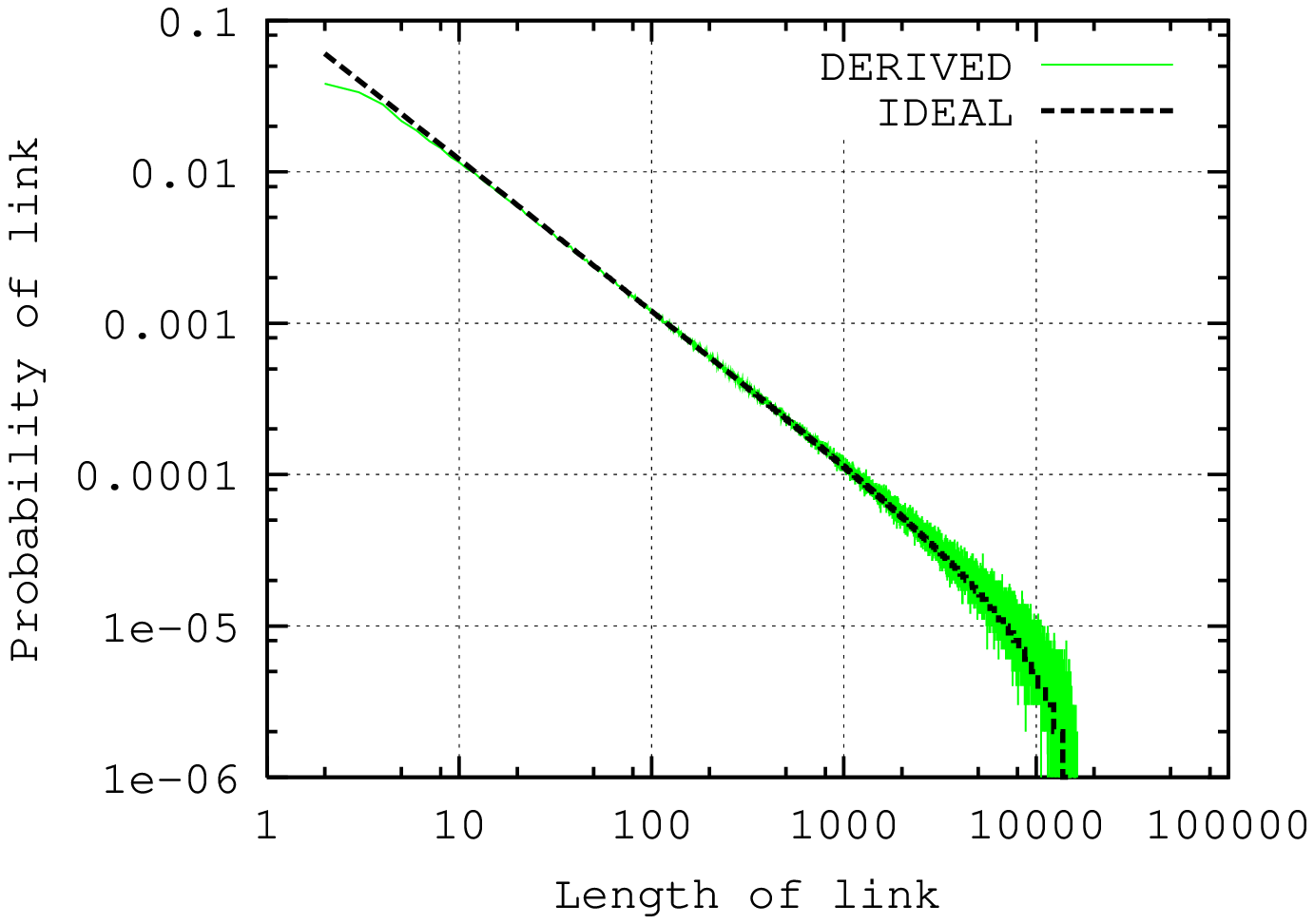, width=0.45\textwidth}}
       \subfigure[Absolute error.
        \label{fig:ERROR}]
       {\epsfig{figure=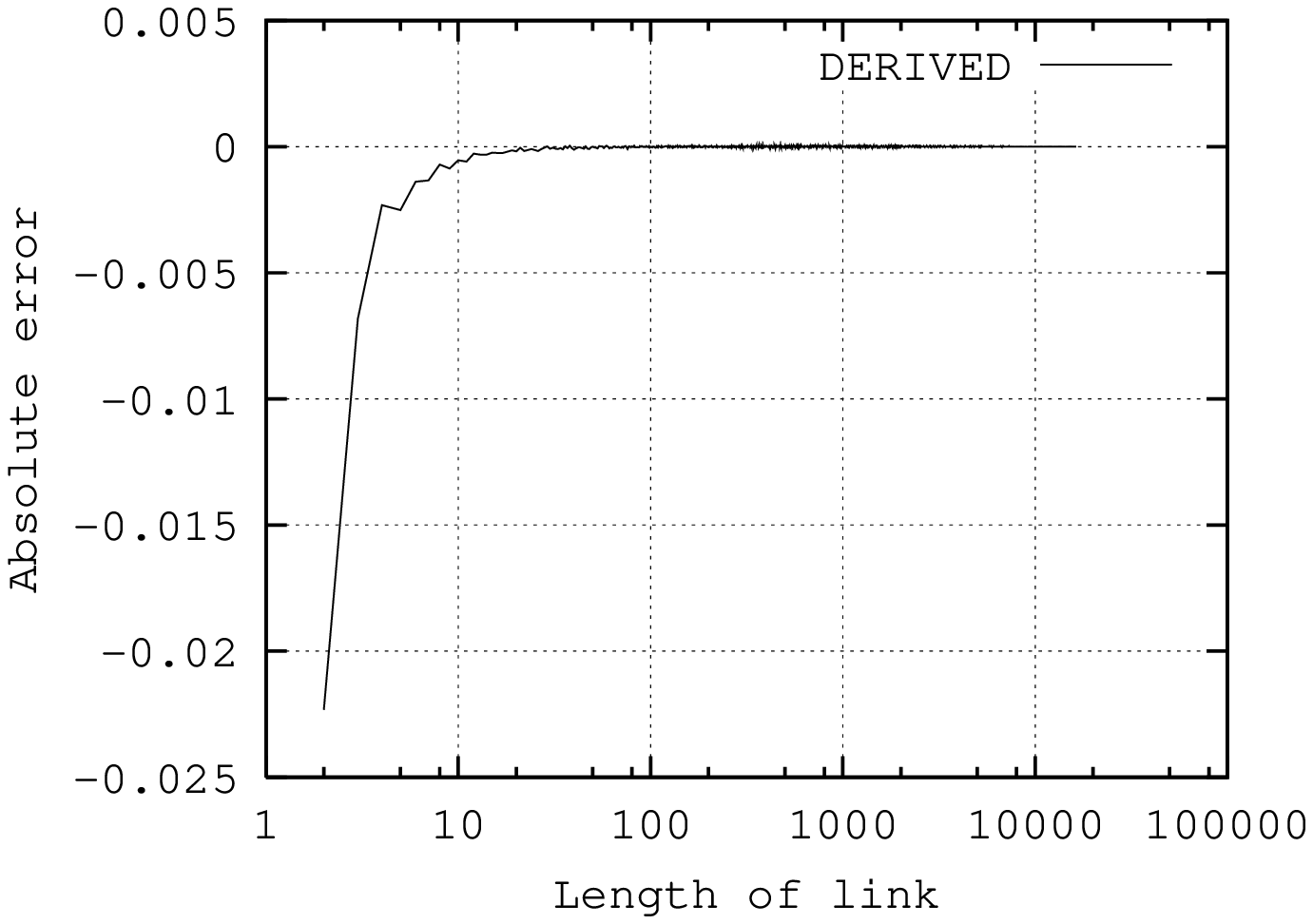, width=0.45\textwidth}}}
\caption{(a) The distribution of long-distance links produced by the 
inverse-distance
heuristic (DERIVED) compared to the ideal inverse power-law distribution with 
exponent $1$ (IDEAL). (b) The absolute error between the derived 
distribution and the ideal inverse power-law distribution with exponent 
$1$.}
\end{figure}

To analyze the performance of the heuristic in practice, we used it to construct a 
network of $2^{14}$ nodes with $14$ links each, ten separate times. After 
averaging the results over the ten networks, we plotted the distribution of 
long-distance links derived from the heuristic, along with the ideal inverse 
power-law distribution with exponent 1, as shown in 
Figure~\ref{fig:DISTRIBUTION}. We see that the derived distribution tracks the 
ideal one very closely, with the largest absolute error being roughly equal to 
$0.022$ for links of length $2$, as shown in the graph of 
Figure~\ref{fig:ERROR}. 

We also performed experiments for an alternative link replacement strategy:
a node chooses its {\em oldest} link to replace with a link to the new node. 
The performance of this strategy is almost as good as the performance 
of our replacement strategy described previously. We omit 
those results because it is difficult to distinguish between the results 
of the two strategies on the scale used for our graphs.

There has also been other related work~\cite{PRU01} on how to construct,
with the support of a central server, random graphs with many desirable 
properties, such as small diameter and guaranteed connectivity with 
high probability. Although it is not clear what kind of fault-tolerance 
properties this approach offers if the central server crashes, or how 
the constructed graph can be used for efficient routing, it is likely 
that similar techniques could be useful in our setting.

\section{Experimental Results}
\label{sec:EXPERIMENTS}

We simulated a network of $n=2^{17}$ nodes at the application level. Each 
node is connected to its immediate neighbors and has $\lg n=17$ long-distance 
links chosen as per the inverse power law distribution with exponent $1$ as 
explained in Section~\ref{sec:UPPERBNDS}. Routing is done greedily by forwarding 
a message to the neighbor closest to its target node. In each simulation, the 
network is set up afresh, and a fraction $p$ of the nodes fail. 
We then repeatedly choose random source and destination nodes that have not 
failed and route a message between them. For each value of 
$p$, we ran $1000$ simulations, delivering $100$ messages
in each simulation, and averaged the number of hops 
for successful searches and the number of failed searches.

With node failures, a node may not be able to find a live
neighbor that is closer to the target node than itself. We studied
three possible strategies to overcome this problem as follows.

\begin{enumerate}
 \item Terminate the search.
 \item Randomly choose another node, deliver the message to 
       this new node and then try to deliver the message from this
       node to the original destination node (similar to 
       the hypercube routing strategy explained in~\cite{LV82}).
 \item Keep track of a fixed number (in our simulations, $5$) 
       of nodes through which the message is last routed and backtrack. 
       When the search reaches a node from where it cannot proceed, it 
       backtracks to the most recently visited node from this list and 
       chooses the next best neighbor to route the message to. 
\end{enumerate}

For all these strategies we note that once a node chooses its best neighbor, 
it does not send the message to any other link if it finds out that the best 
neighbor has failed.

\begin{figure}
\centering
  \mbox{\subfigure[Fraction of failed searches.]
       {\epsfig{figure=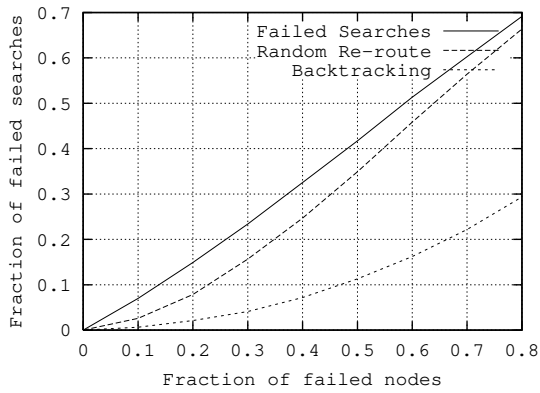, width=0.45\textwidth}}\quad
       \subfigure[Average delivery time for successful searches.]
       {\epsfig{figure=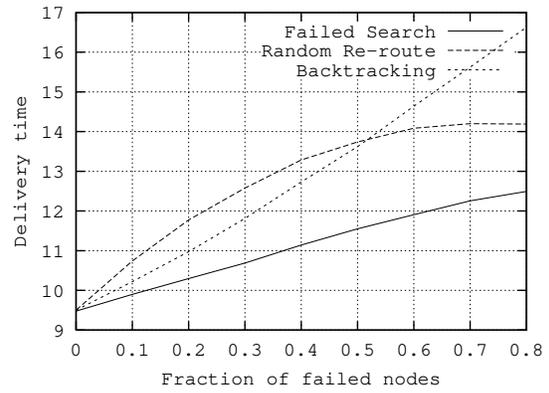, width=0.45\textwidth}}}
\caption{(a) The fraction of messages that fail to be delivered
as a function of the fraction of failed nodes. (b) The average delivery
time for successful searches as a function of the fraction of
failed nodes.}
\label{fig:RESULTS}
\end{figure}

Figure~\ref{fig:RESULTS} shows the fraction of messages that 
fail to be delivered and the number of hops for successful 
searches versus the fraction of failed nodes. We see that the 
system behaves well even with a large number of failed nodes.
In addition, backtracking 
gives a significant improvement in reducing the number of failures as 
compared to the other two methods, although it may take a longer time 
for delivery. We see that in the case of random rerouting,
the average delivery time does not increase too much as the probability 
of node failure increases.
This happens because quite a few of the searches fail, so the ones
that succeed (with a few hops) lead to a small average delivery time.

Our results may not be directly comparable to those of CAN\cite{SR01}
and Chord\cite{CH01}, since they use different simulators for 
their experiments. However, to the extent that the results are comparable, 
our methods appear to perform as well as theirs.
Even if we just terminate the search, we get less than $p$ fraction
of failed searches with $p$ fraction of failed nodes. Chord\cite{CH01} 
has roughly the same performance {\em after} their network stabilizes
using their repair mechanism. Further, with backtracking we see that with
$80\%$ failed nodes, we still get less than $30\%$ failed searches.
These results are very promising and it would be interesting to 
study backtracking analytically.

\begin{center}
\begin{figure}[ht]
  \centerline{\mbox{\epsfig{figure=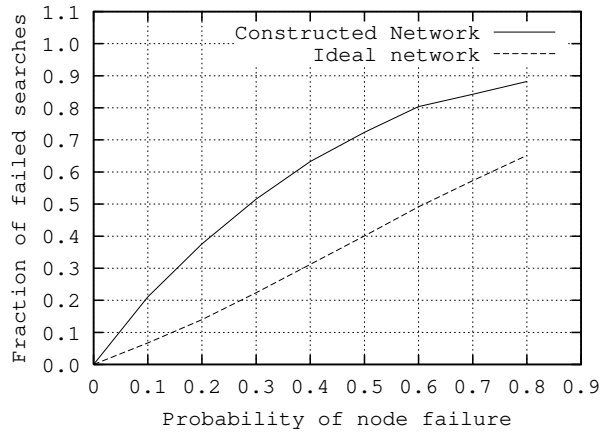, width=0.5\textwidth}}}
  \caption{Fraction of failed searches.}
\label{fig:COMPARE}
\end{figure}
\end{center}

We also compared the performance of the ideal network and that of the
network constructed using the heuristics given in Section~\ref{sec:RANDOMGRAPHS}.
We ran $10$ iterations of constructing a network of $16384$ nodes, both
ideally as well as according to the heuristic, and delivered $1000$ messages
between randomly chosen nodes.
Figure~\ref{fig:COMPARE} shows the number of failed searches as the probability 
of node failure increases. We see that although the network constructed
using the heuristic does not perform as well as the ideal network, the
number of failed searches is comparable. 

\section{Conclusions and Future Work}
\label{sec:CONCLUSIONS}

\begin{table}[ht]
\begin{center}
\begin{tabular}{|c|c|c|c|}

\hline
Model&
Number of Links $\ell$
&Upper Bound 
&Lower Bound\\

\hline
\multirow{3}*{No failures}
&1\bigstrut
&$O(\log^2 n)$\bigstrut
&$\Omega(\frac{\log^2 n}{\log \log n})$\bigstrut\\

&$[1, \lg n]$\bigstrut
&$O(\frac{\log^2 n}{\ell})$\bigstrut
&$\Omega(\frac{\log^2 n}{\ell \log \log n})$\bigstrut\\

&$[\lg n, n^c]$\bigstrut
&$O(\frac{\log n}{\log b})$\bigstrut
&$\Omega(\frac{\log n}{\log \ell})$\bigstrut\\

\hline
\hline

\multirow{2}*{Pr[Link present]=$p$}
&$[1, \lg n]$\bigstrut
&$O(\frac{\log^2 n}{p\ell})$\bigstrut
&-\bigstrut\\

&$[\lg n, n^c]$\bigstrut
&$O(\frac{b\log n}{p})$\bigstrut
&-\bigstrut\\

\hline
\hline

\multirow{2}*{Pr[Node present]=$p$}
&\multirow{2}*{$[1, \lg n]$}
&\multirow{2}*{$O(\frac{\log^2 n}{p\ell})$}
&\multirow{2}*{-}\\

&&&\\

\hline
\end{tabular}
\end{center}
\caption{Summary of upper and lower bounds for routing.\protect\footnotemark}
\label{table-results}
\end{table}
\footnotetext{In the upper bound with 
$(\lg n, n^c]$ links, the number of links
$\ell=O(b\log_b n)$. Also, the deterministic strategy 
used for links $\ell \in (\lg n, n^c]$, 
with link failures is 
slightly different that the one with no failures, 
and $\ell=O(\log_b n)$.
In the lower bound column, the bound for $[1,\lg n]$ links is for
one-sided routing.}

Table~\ref{table-results} summarizes our upper and lower bounds.
We have shown that greedy routing in an overlay network organized as a
random graph in a metric space can be a nearly optimal mechanism for
searching in a peer-to-peer system, even in the presence of
many faults.  We see this as an important first step in the design of
efficient algorithms for such networks, but many issues still need to
be addressed.  Our results mostly apply to one-dimensional metric
spaces like the line or a circle.  One interesting possibility is
whether similar strategies would work for higher-dimensional spaces,
particularly ones in which some of the dimensions represent the actual
physical distribution of the nodes in real space; good
network-building and search mechanisms for this model might allow
efficient location of nearby instances of a resource without having to
resort to local flooding (as in~\cite{KKD01}).
Another promising direction would be to study the security properties
of greedy routing schemes to see how they can be adapted to provide
desirable properties like anonymity or robustness against Byzantine
failures.

\section{Acknowledgments}

The authors are grateful to Ben Reichardt for pointing out an error in
an earlier version of Lemma~\ref{lemma-log-drop}.

\bibliographystyle{abbrv}
\bibliography{paper}

\end{document}